\documentclass[amsfonts,amssymb,amsmath,floatfix,a4paper,superscriptaddress,nofootinbib,preprint,prd]{revtex4}
\usepackage{graphicx}
\usepackage{axodraw}
\usepackage{rotating}

\begin{document}
\preprint{\vbox{\hbox{ADP-02-74-T513}
	\hbox{JLAB-THY-02-26}}}

\title{Moments of Isovector Quark
    Distributions from Lattice QCD}
\author{W. Detmold}
\affiliation{Special Research Centre for the Subatomic Structure of
        Matter and Department of Physics and Mathematical Physics,
        University of Adelaide, SA 5005, Australia}
\author{W. Melnitchouk}
\affiliation{Jefferson Lab, 12000 Jefferson Avenue,
        Newport News, VA 23606}
\author{A. W. Thomas}
\affiliation{Special Research Centre for the Subatomic Structure of
        Matter and Department of Physics and Mathematical Physics,
        University of Adelaide, SA 5005, Australia}


\begin{abstract}
We present a complete analysis of the chiral extrapolation of lattice
moments of all twist-2 isovector quark distributions, including
corrections from $N\pi$ and $\Delta \pi$ loops.
Even though the $\Delta$ resonance formally gives rise to higher order
non-analytic structure, the coefficients of the higher order terms for
the helicity and transversity moments are large and cancel much of the
curvature generated by the wave function renormalization.
The net effect is that, whereas the unpolarized moments exhibit
considerable curvature, the polarized moments show little deviation from
linearity as the chiral limit is approached.
\end{abstract}

\maketitle


\section{Introduction}
\label{S:introduction}

Resolving the quark and gluon structure of the nucleon remains one of
the central challenges in strong interaction physics \cite{Thomas:kw}.
Information about the nucleon's internal structure is parameterized in
the form of leading twist parton distribution functions (PDFs),
which are interpreted as probability distributions for finding specific
partons (quarks, antiquarks, gluons) in the nucleon in the infinite
momentum frame.
PDFs have been measured in a variety of high energy processes ranging
from deep-inelastic lepton scattering to Drell-Yan and massive vector
boson production in hadron--hadron collisions.
A wealth of experimental information now exists on spin-averaged PDFs,
and an increasing amount of data is being accumulated on spin-dependent
PDFs \cite{DATAREVIEW}.

The fact that such a vast array of high energy data can be analyzed in
terms of a universal set of PDFs stems from the factorization property
of high energy scattering processes, in which the short and long
distance components of scattering amplitudes can be separated
according to a well-defined procedure.
Factorization theorems allow a given differential cross section, or
structure function, $F$, to be written (as a function of the light-cone
momentum fraction $x$ at a scale $Q^2$) in terms of a convolution of hard,
perturbatively calculable coefficient functions, $C_i$, with the PDFs,
$f_i$, describing the soft, non-perturbative physics \cite{FACT_THMS}:
\begin{eqnarray}
F(x,Q^2) &=&
\sum_i \int dz\ C_i(x/z,Q^2/\mu^2,\alpha_s(\mu^2))\
f_i(z,\alpha_s(\mu^2))\ ,
\end{eqnarray}
where $\mu$ is the factorization scale.
The coefficient functions are scale and process dependent, while the PDFs
are process independent, and hence can be used to parameterize a wide
variety of high energy data.

Because the PDFs cannot be calculated within perturbative QCD, the
approach commonly used in global analyses of high energy data is to
simply parameterize the PDFs, without attempting to assess their
dynamical origin \cite{CTEQ,MRS,GRV,BB}.
Once fitted at a particular scale, they can be evolved to any other
scale through the DGLAP $Q^2$-evolution equations \cite{DGLAP}.
The focus in this approach is not so much on understanding the
non-perturbative (confinement) physics responsible for the specific
features of the PDFs, but rather on understanding the higher order QCD
corrections for high energy processes.

In a more ambitious approach one would like to extract information about
non-perturbative hadron structure from the PDFs.
However, without an analytic solution of QCD in the low energy realm one
must rely to varying degrees on models of (or approximations to) QCD
within which to interpret the data.
An extensive phenomenology has been developed over the years within
studies of QCD-motivated models, and in some cases remarkable predictions
have been made from the insight gained into the non-perturbative structure
of the nucleon.
An example is the $\bar d-\bar u$ asymmetry, predicted \cite{AWT83} on
the basis of the nucleon's pion cloud \cite{EARLY}, which has been
spectacularly confirmed in recent experiments at CERN, Fermilab and DESY
\cite{EXPT}.
Other predictions, such as asymmetries between strange and anti-strange
\cite{STRANGE}, and spin-dependent sea quark distributions,
$\Delta \bar u - \Delta \bar d$ \cite{POLSEA}, still await definitive
experimental confirmation.
Note that none of these could be anticipated without insight into the
non-perturbative structure of QCD.

Despite the phenomenological successes in correlating deep-inelastic
and other high energy data with low energy hadron structure, the
{\em ad hoc} nature of some of the assumptions made in deriving the low
energy models from QCD leaves open questions about the ability to
reliably assign systematic errors to the model predictions.
One approach in which structure functions can be studied
systematically from first principles, and which at the same time
allows one to search for and identify the relevant low energy QCD
degrees of freedom, is lattice QCD.

Lattice QCD is rapidly developing into an extremely useful and
practical tool with which to study hadronic structure \cite{KFLIU}.
There, the equations of motion are solved numerically by discretizing
space-time into a lattice, with quarks occupying the lattice sites and
gluons represented by the links between the sites.
Meaningful numerical results can be obtained by Wick rotating the QCD
action into Euclidean space.
However, because the leading twist PDFs are light-cone correlation
functions (involving currents with space-time separation,
$z^2 - (ct)^2 \approx 0$), it is, in practice, not possible to calculate
PDFs directly in Euclidean space --- in this case a null vector would
require each space-time component to approach zero.
Instead, one uses the operator product expansion to formally express the
moments of the PDFs in terms of hadronic matrix elements of local
operators, which can then be calculated numerically.

In relating the lattice moments to experiment, a number of
extrapolations must be performed.  Since lattice calculations are
performed at some finite lattice spacing, $a$, the results must be
extrapolated to the continuum limit, $a \to 0$, which can be done by
calculating at two or more values of $a$.  Furthermore, finite volume
effects associated with the size of the lattice must be controlled ---
working with a volume that is too small can result in the omission of
important physics, arising from the long-range part of the nucleon
wave function.  Finally, since current lattice simulations typically
use quark masses $m_q^{\rm latt}$ above 30~MeV, an extrapolation to
physical masses, $m_q^{\rm phys} \approx 5$~MeV, is necessary.
Earlier work on moments of spin-averaged PDFs
\cite{Detmold:2001jb,Detmold:2001dv} found that whereas the lattice
calculations yielded results typically 50\% larger than experiment
when extrapolated linearly to $m_q^{\rm phys}$, inclusion of the
nonlinear, non-analytic dependence on $m_q$ arising from the long
range structure of the nucleon removes most of the discrepancy.

In this paper we extend the analysis to the polarized sector, which is
important for several reasons.  Firstly, for many years lattice
calculations of the axial vector charge, $g_A$, have tended to lie
10\% or more below the experimental value determined from neutron
$\beta$ decay.  Since this represents one of the benchmark
calculations in lattice QCD, it is vital that the source of this
discrepancy be identified.  A preliminary analysis of the effects of
chiral loops found \cite{WDThesis,CHINA} that the inclusion of the
leading non-analytic (LNA) behavior associated with $\pi N$
intermediate states in the extrapolation of $g_A$ to $m_q^{\rm phys}$
decreased the value of $g_A$, thereby making the disagreement worse.
On the other hand, one knows that the $\Delta$ resonance plays an
important role in hadronic physics, so a more thorough investigation of
its effects on spin-dependent PDFs is necessary before definitive
conclusions can be drawn.
Indeed, we find that although the $\Delta$ contributions formally enter
at higher order in $m_\pi$, the coefficients of these next-to-leading
non-analytic terms are large, and their effects cannot be ignored in any
quantitative analysis. 
In addition, since there are currently no data at all on the transversity
distribution in the nucleon, lattice calculations of several low 
transversity moments provide predictions which can be tested by
future measurements.  In order to make these predictions
reliable, it is essential that the lattice calculations be reanalyzed
to take into account the chiral corrections entering extrapolations to
$m_q^{\rm phys}$.

The remainder of this manuscript is structured in the following manner.
In Section~\ref{S:lattice} we describe the calculations of the moments
of PDFs from matrix elements of local operators, and summarize the
details of extant lattice calculations.
In Section~\ref{S:chiral} we first examine the constraints from chiral
perturbation theory and the heavy quark limit on the behavior of the
moments of the various distributions as a function of the quark mass.
The importance of higher order terms in the chiral expansion is then
investigated within a model which preserves the non-analytic
behavior of chiral perturbation theory.
This information is used to construct effective parameterizations of the
quark mass dependence of these moments, which are then used to extrapolate
the available lattice data in Section~\ref{S:xtrap}.
Finally, in Section~\ref{S:conclusion} we discuss the results of this
analysis and draw conclusions.

\section{Lattice Moments of Parton Distribution Functions}
\label{S:lattice}

\subsection{Definitions}

The moments of the spin-independent, $q=q^\uparrow + q^\downarrow$,
helicity, $\Delta q=q^\uparrow - q^\downarrow$, and transversity,
$\delta q=q^{\top} - q^\perp$, distributions are defined as:
\begin{subequations}
\label{E:moments}
\begin{eqnarray} 
  \langle x^n \rangle_{q}&=&\int_0^1 dx\,
  x^n\ [q(x)-(-1)^n \bar q(x)]\ ,  \\
  \langle x^n \rangle_{\Delta q} &=&\int_0^1 dx\,
  x^n\ [\Delta q(x)+(-1)^n \Delta \bar  q(x)]\ , \\
  \langle x^n \rangle_{\delta q} &=&\int_0^1 dx\, 
  x^n\ [\delta q(x)-(-1)^n \delta \bar q(x)] \,,
\end{eqnarray}
\end{subequations}%
where $q^{\uparrow(\downarrow)}$ corresponds to quarks with helicity
aligned (anti-aligned) with that of a longitudinally polarized target,
and $q^{\top(\perp)}$ corresponds to quarks with spin aligned
(anti-aligned) with that of transversely polarized
target\footnote{Note that from their definition,
Eqs.~(\ref{E:moments}), the moments alternate between the total
($q+\overline q$) and valence ($q-\overline q$) distributions,
depending on whether $n$ is even or odd.}.  At leading twist, these
moments depend on ground state nucleon matrix elements of the
operators
\begin{subequations}
  \label{E:operators}
\begin{eqnarray}
  {\cal O}^{\mu_{1}\ldots\mu_{n}}_{q} &=&
  i^{n-1}\ \overline{\psi}\
  \gamma^{\{\mu_{1}}\, D^{\mu_{2}}\cdots \,D^{\mu_{n}\}}\psi   \,,
  \\
  {\cal O}^{\mu_{1}\ldots\mu_{n}}_{\Delta q} &=&
  i^{n-1}\ \overline{\psi}\
  \gamma^{\{\mu_1}\,\gamma_{5}\, D^{\mu_{2}}\cdots
  \,D^{\mu_{n}\}}\psi \,,
  \\
  {\cal O}^{\alpha\mu_{1}\ldots\mu_{n}}_{\delta q} &=&
  i^{n-1}\ \overline{\psi}\
  \sigma^{\alpha\{\mu_1}\,\gamma_{5}\, D^{\mu_{2}}\cdots
  \,D^{\mu_{n}\}}\psi \,,
\end{eqnarray}
\end{subequations}%
respectively.  Thus, for a nucleon of mass $M$, momentum $P$, and spin
$S$, one has:
\begin{subequations}
  \label{E:matrixelts}
\begin{eqnarray}
  \langle P,S |{\cal
  O}^{\mu_{1}\ldots\mu_{n}}_{q}|P,S \rangle &=& 2\ \langle x^{n-1}
  \rangle_{q}\ P^{\{\mu_{1}}\cdots P^{\mu_{n}\}} - \text{traces} \,,
  \\
  \langle P,S |{\cal O}^{\mu_{1}\ldots\mu_{n}}_{\Delta q}|P,S
  \rangle &=& 2\ \langle x^{n-1} \rangle_{\Delta
  q}\ M\ S^{\{\mu_1}P^{\mu_{2}}\cdots P^{\mu_{n}\}} - \text{traces} \,,
  \\
  \langle P,S |{\cal O}^{\alpha\mu_{1}\ldots\mu_{n}}_{\delta q}|P,S \rangle
  &=& 2\ \langle x^{n-1} \rangle_{\delta q}\
  M\ S^{[\alpha}P^{\{\mu_1]}P^{\mu_{2}}\cdots P^{\mu_{n}\}} -
  \text{traces} \,,
\end{eqnarray}
\end{subequations}%
where the braces, \{ $\cdots$ \} ([ $\cdots$ ]), imply symmetrization
(anti-symmetrization) of indices, and the `traces' (containing
contractions $g^{\mu_i\mu_j}$, etc.) are subtracted to make the matrix
elements traceless in order that they transform irreducibly.
At higher twist (suppressed by powers of $1/Q^2$), more complicated
operators involving both quark and gluon fields contribute.

\subsection{Lattice Operators}

The construction of the relations (\ref{E:matrixelts}) between moments of
PDFs and matrix elements of local operators relies on the symmetry group
of the Euclidean space in which one works.
When formulated on a discrete space-time lattice, the symmetry group is
reduced and the discretized implementation of these operators introduces
several technical complications.

The discrete nature of the lattice topology means that the symmetry
group of the Euclidean continuum, the orthogonal group O(4), is broken
to its hyper-cubic subgroup, H(4) (the group of 192 discrete rotations
which map the lattice onto itself) \cite{DolgovThesis}.
Unfortunately, operators in irreducible representations of O(4) may
transform reducibly under H(4) and this may result in mixing with
operators from lower dimensional multiplets under renormalization.
Consequently, care must be exercised in the choice of operators
on the lattice.
For example, the continuum operator ${\cal O}^{\mu\nu}_{q}$, which
corresponds to the momentum carried by quarks, can be represented
on the lattice by either
${\cal O}_q^{(a)} = \overline{\psi}\gamma^{\{1}D^{4\}}\psi$
(belonging to a ${\bf 6}$ representation) or
${\cal O}_q^{(b)}
= \overline{\psi}\gamma^{4}D^{4}\psi -\frac{1}{3} \sum_{i=1}^{3}
  \overline{\psi}\gamma^{i}D^{i}\psi$
(belonging to a ${\bf 3}$ representation).
This may be regarded as an advantage since in the $a\to0$ limit these
operators are identical and any difference at non-zero lattice spacing
allows for an estimate of the remaining finite size lattice artifacts to
be made.
In practice, this is currently somewhat ambitious, as the operator
${\cal O}_q^{(a)}$ requires that the hadron source should have non-zero
momentum components, which leads to a statistically less well determined
result.
Consequently, for the operator ${\cal O}^{\mu\nu}_{q}$ we retain only the
data corresponding to ${\cal O}_q^{(b)}$.
The operators associated with the unpolarized $n=2$ and 3 moments are
given by
$ {\cal O}_q^{ \{114 \} }
- \frac{1}{2} \sum_{i=2}^3 {\cal O}_q^{ \{ ii4 \} }$
and
$ {\cal O}_q^{ \{ 1144 \} } + {\cal O}_q^{ \{ 2233 \} }
- {\cal O}_q^{ \{ 1133 \} } - {\cal O}_q^{ \{ 2244 \} }$, respectively.

For the spin-dependent moments, the operator corresponding to the axial
charge is given by
${\cal O}_{\Delta q}^3 = \overline{\psi} \gamma^5 \gamma^3 \psi$.  
However, for the $n=1$ moment one can have on the lattice either
${\cal O}_{\Delta q}^{(a)}
= \overline{\psi} \gamma^5 \gamma^{\{1}D^{3\}} \psi$ or
${\cal O}_{\Delta q}^{(b)}
= \overline{\psi} \gamma^5 \gamma^{\{3}D^{4\}} \psi$.
Once again, since the operator ${\cal O}_{\Delta q}^{(a)}$ requires
non-zero momentum, we shall keep only the data corresponding to the
better determined ${\cal O}_{\Delta q}^{(b)}$ operator.
The operators required to calculate other moments in
Eqs.~(\ref{E:moments}) are described in Ref.~\cite{Dolgov:2002zm}.

For spin greater than 3, there are no unique, irreducible representations
in H(4) for the twist-2 operators.
This means that the operators for moments $n > 3$ will \emph{inevitably}
mix with lower dimensional (or lower spin) operators.
To unambiguously extract information about these moments, one would need
to consider all representations for a given spin, and, with sufficiently
accurate data, deduce the matrix elements of the high spin operators from
the low spin operators with which they mix.
Because of these difficulties, all lattice calculations have so far been
restricted to moments with $n \leq 3$.
Nevertheless, some features of the PDFs can be reasonably reconstructed
from just the lowest few moments, as described in
Ref.~\cite{Detmold:2001dv}.

Further subtleties arise when we consider the non-perturbative
renormalization of these operators and their matching to other
renormalization schemes.
An operator, ${\cal O}_{\rm latt}(a)$, calculated using the lattice
regularization scheme, is connected to other schemes, for example
$\overline{\rm MS}$, by a renormalization factor:
\begin{displaymath}
    {\cal O}_{\overline{\rm MS}}(\mu)=Z_{\cal O}(\mu,a)\ {\cal 
    O}_{\rm latt}(a)\ ,
\end{displaymath}
where $\mu$ is the renormalization scale.
To provide results in standard schemes, the renormalization functions,
$Z_{\cal O}$, must therefore be calculated for each operator used.
While this is done perturbatively in most calculations, 
non-perturbative determinations now exist \cite{Gockeler:1998rw}.
In what follows, results are presented in the $\overline{\rm MS}$ scheme
at a renormalization scale $\mu^2\approx 4$~GeV$^2$.

\subsection{Lattice Calculations}

The first calculations of structure functions within lattice QCD were
performed in the late 1980s by Martinelli and Sachrajda.  Their
pioneering calculations of quark distributions of the pion
\cite{Martinelli:1987zd} and nucleon \cite{Martinelli:1989xs} were
ambitious, given the speed of the computers available at the time.
More recently, various calculations of greater precision have been
performed \cite{Gusken:1989ad, Dong:1995rx, Liu:1994ab,
Fukugita:1995fh, Aoki:1997pi, Gusken:1999as,Gockeler:1996wg,
Gockeler:1997jk, BEST, SchierholzPC, Dolgov:2000ca, DolgovThesis,
Dolgov:2002zm, Gupta:1991sn, Sasaki:2001th, SasakiPC}.  In the present
analysis we will focus mainly on the more recent
QCDSF~\cite{Gockeler:1996wg, Gockeler:1997jk, BEST,
SchierholzPC} and LHPC-SESAM simulations~\cite{Dolgov:2000ca,
DolgovThesis, Dolgov:2002zm}.
The older data sets from Gupta \emph{et al.} \cite{Gupta:1991sn} have
large uncertainties associated with renormalization, while the
statistical precision of Refs.~\cite{Dong:1995rx,Liu:1994ab} is
comparatively low.
In addition, several groups (notably the KEK
\cite{Fukugita:1995fh, Aoki:1997pi}, Riken-BNL-Columbia (RBC)
\cite{Sasaki:2001th,SasakiPC} and SESAM \cite{Gusken:1999as} collaborations)
have put particular emphasis on the $n=0$ moments of the helicity and
transversity distributions --- the axial and tensor charges.
The simulations have been made using various quark and gluon actions,
on different lattices and at different couplings.  They have been
performed primarily in the quenched approximation, although more
recently the LHPC \cite{Dolgov:2002zm} and UKQCD/QCDSF
\cite{SchierholzPC} groups have begun to investigate the effects of
unquenching.  In Table~\ref{T:data} we summarize the data used here,
for which PDF moments and the corresponding pion masses are published.

Before including the data sets in our analysis, we impose a simple cut
to reduce finite volume effects.
In lattice calculations of any observable, one must ensure that the
lattice size is large enough for results not to be dependent on the
(unphysical) boundary conditions.
This applies particularly to calculations involving low energy states
such as the nucleon where the effects of the pion cloud are known to
be especially important.
Being the lightest, and therefore longest range, asymptotic correlation of
quarks and gluons, pions are most sensitive to the boundary conditions.
To avoid these difficulties, we require that the lattice volume is large
enough that a pion will fit comfortably within it without ``feeling the
edges of the box''.
A pion of mass $m_{\pi}$ has a corresponding Compton wavelength of
order $\lambda_{\pi}\sim1/m_{\pi}$, and, to avoid interference of the
pion with its periodic copies, we require that the smallest dimension
of the lattice box ($L$) satisfies the constraint
\begin{equation}
    L\, >\, 4\,\lambda_{\pi}\sim\frac{4}{m_{\pi}}\ .
\end{equation}
The factor of $4$ in this formula is popular \cite{DolgovThesis}, although
somewhat arbitrary.
This argument indicates that the lowest mass data point of
Ref.~\cite{Gockeler:1997jk} and the lightest unquenched points from
Ref.~\cite{Dolgov:2002zm} should be excluded from the analysis.
\begin{sidewaystable}[!ht]
    \begin{center}
    \begin{ruledtabular}
        \begin{tabular}{cccccccc}
        Reference & Q/U &  Quark Action & Lattice & $a$ [fm]  & $m_\pi$ [GeV]
        & Moments & Symbol
        \\
        \hline\hline
        QCDSF \cite{Gockeler:1996wg} & Q & Wilson & $16^{3}\times32$
        & 0.1 & 0.6 -- 1.0 & All & {\scriptsize $\blacktriangle$}
        \\
        QCDSF \cite{Gockeler:1997jk} & Q & Wilson & $24^{3}\times32$
        & 0.1 & 0.35 -- 0.6 & All & {\tiny $\blacksquare$}
        \\
        QCDSF \cite{BEST}& Q & NPIC & $16^{3}\times32$  &  0.1 & 0.6
        -- 1.0 & All & $\times$
        \\
        QCDSF \cite{SchierholzPC} & Q & NPIC & $16^{3}\times32$  & 0.1
        & 0.65 -- 1.2 & $\langle 1\rangle_{\Delta q}$, $\langle
        x^2\rangle_{\Delta q}$ & {\tiny $\blacklozenge$} 
        \\
        & Q & NPIC & $24^{3}\times48$  &  0.075 
        & 0.7 -- 1.2 & $\langle 1\rangle_{\Delta q}$, $\langle
        x^2\rangle_{\Delta q}$ & $\bullet$ 
        \\
        & Q & NPIC & $32^{3}\times48$  &  0.05 
        & 0.6 -- 1.25 & $\langle 1\rangle_{\Delta q}$, $\langle
        x^2\rangle_{\Delta q}$ & $\times\hspace{-1.81ex}+$ 
        \\
        MIT \cite{Dolgov:2002zm} & Q & Wilson & $16^{3}\times32$ & 0.1
        & 0.58 -- 0.82 & All & {\scriptsize $\bigstar$}
        \\
        MIT-SESAM \cite{Dolgov:2002zm} & U & Wilson  &
        $16^{3}\times32$ & 0.1& 0.63 -- 1.0 & All & {\tiny $\lozenge$}
        \\
        MIT-SCRI \cite{Dolgov:2002zm} &  U & Wilson  &
        $16^{3}\times32$ & 0.1& 0.48 -- 0.67 & All & {\tiny $\square$}
        \\
        KEK \cite{Fukugita:1995fh,Aoki:1997pi} & Q & Wilson & $16^{3}\times20$ &
        0.14 & 0.52 -- 0.97 & $\langle
        1\rangle_{\Delta q}$, $\langle 1\rangle_{\delta q}$ &
        {\scriptsize $\blacktriangledown$}
        \end{tabular}
    \end{ruledtabular}
    \caption{Simulation parameters for lattice calculations of the 
        moments of PDFs included in our analysis.
        Q/U corresponds to quenched/unquenched simulations, and NPIC
        denotes the nonperturbatively improved clover quark action.
        ``All'' moments correspond to $\langle x^i\rangle_q$ for
        $i=1,2,3$, $\langle x^i\rangle_{\Delta q}$ for $i=0,1,2$, and
        $\langle x^i\rangle_{\delta q}$ for $i=0,1$.
        The symbols shown in the final column correspond to those
        plotted in Figs.~\ref{F:unpolarized}, \ref{F:polarized} and
        \ref{F:transversity}.}
      \label{T:data}
\end{center}
\end{sidewaystable}

\begin{figure}[!ht]
  \begin{center}
      \includegraphics[width=12cm]{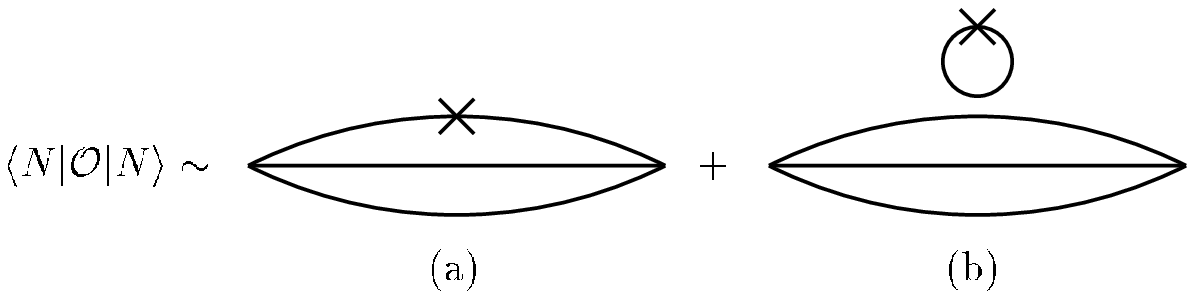}
      \caption{Connected (a) and disconnected (b) contributions to the matrix 
        elements of an operator (indicated by the cross). Such
        diagrams occur in quenched QCD as well as in full QCD.}
      \label{F:disc}
  \end{center}
\end{figure}

In terms of quark flow, the evaluation of matrix elements of the
operators in Eqs.~(\ref{E:operators})
includes both connected and disconnected diagrams, corresponding to
operator insertions in quark lines which are connected or disconnected
(except through gluon lines) with the nucleon source --- see
Fig.~\ref{F:disc}.
Since the numerical computation of disconnected diagrams is
considerably more difficult, only exploratory studies of these have
thus far been completed \cite{Gusken:1999as}, and the data analyzed
here include only connected contributions.  However, because the
disconnected contributions are flavor independent (for equal $u$ and
$d$ quark masses), they exactly cancel in the {\em difference} of $u$
and $d$ moments.  Therefore, until more complete lattice simulations
become available, one can only compare lattice moments of the flavor
non-singlet $u-d$ distribution with moments of phenomenological PDFs
\cite{CTEQ,MRS,GRV}.

Whilst the chiral behavior of quenched QCD is different from that of
full QCD \cite{Chen:2001gr}, in the region where current lattice data
from both quenched and unquenched simulations are available, the
differences are well within the statistical errors, indicating that
internal quark loops do not play a significant role over this mass
range.  As calculations begin to probe lighter quark masses, the
differences should become more apparent and it will become necessary
to analyze quenched and unquenched data separately
\cite{Young:2001nc}.  Until the differences become statistically
distinguishable, however, we shall combine the data from the two sets of
simulations.

\section{Chiral Behavior of PDF Moments}
\label{S:chiral}

To compare the lattice results with the experimentally measured moments,
one must extrapolate the data from the lowest quark mass used
($\sim 50$~MeV) to the physical value ($\sim$ 5~MeV).
This is commonly done by assuming that the moments depend linearly on the
quark mass.
However, as discussed in Ref.~\cite{Detmold:2001jb}, such a linear
extrapolation overestimates the experimental values of the unpolarized
isovector moments \cite{CTEQ,MRS,GRV} by some 50\% in all cases.
Since the discrepancy persists in unquenched simulations
\cite{Dolgov:2000ca,DolgovThesis,Dolgov:2002zm}, it suggests that
important physics is being omitted from either the lattice calculations
or their extrapolations.
In Refs.~\cite{Detmold:2001jb,Detmold:2001dv} the chiral behavior of
the moments of the unpolarized isovector distributions was found to be
vital in resolving this discrepancy.
Here we summarize the results of the earlier unpolarized study, and
extend the analysis to the moments of the spin-dependent isovector
distributions in the nucleon.

\subsection{Chiral Symmetry and Leading Non-analytic Behavior}

The spontaneous breaking of the chiral SU$_L$(N$_f$)$\times$SU$_R$(N$_f$)
symmetry of QCD generates the nearly massless Goldstone bosons (pions),
whose importance in hadron structure is well documented.
At small pion masses, hadronic observables can be systematically expanded
in a series in $m_\pi$ --- chiral perturbation theory ($\chi$PT) \cite{CHIPT}.
The expansion coefficients are generally free parameters, determined from
phenomenology.
One of the unique consequences of pion loops, however, is the appearance
of non-analytic behavior in the quark mass.
{}From the Gell-Mann--Oakes--Renner relation one finds that
$m_\pi^2 \sim m_q$ at small $m_\pi$, so that terms involving odd powers
or logarithms of $m_\pi$ are non-analytic functions of the quark mass.
Their presence can lead to highly nonlinear behavior near the chiral
limit ($m_\pi\to0$) \cite{Detmold:2001hq}.
Because the non-analytic terms arise from the infrared behavior of the
chiral loops, they are generally model independent.

The leading order (in $m_\pi$) non-analytic term in the expansion of the
moments of PDFs was shown in Ref.~\cite{TMS} to have the generic behavior
$m_\pi^2 \log m_\pi$ arising from $\pi N$ intermediate states.
This was later confirmed in $\chi$PT \cite{CHPT}, where the coefficients
of these terms were also calculated.
In Ref.~\cite{Detmold:2001jb} a low order chiral expansion for the moments
of the non-singlet distribution, $u-d$, was developed, incorporating the
LNA behavior of the moments as a function of $m_q$ and also connecting to
the heavy quark limit (in which quark distributions become
$\delta$-functions centered at $x=1/3$) \cite{Detmold:2001dv}.
For the moments of the unpolarized isovector distribution, these
considerations lead to the following functional form for the moments
\cite{Detmold:2001dv}:
\begin{equation}
\label{xtrap}
       \langle x^n\rangle_{u-d}\ =\ a_n \left( 1 + c_{\rm
       LNA}m_\pi^2\log\frac{m_\pi^2}{m_\pi^2+\mu^2} \right)\ +\
       b_n\frac{m_\pi^2}{m_\pi^2+m_{b,n}^2}\ ,
\end{equation}
where (for $n>0$) the chiral coefficient
$c_{\rm LNA} = -(1 + 3 g_A^2)/(4\pi f_\pi)^2$ \cite{CHPT},
and $b_n$ is a constant constrained by the heavy quark limit:
\begin{equation}
  b_n = \frac{1}{3^n} - a_n \left( 1 - \mu^2 c_{\rm LNA} \right)\ .
\end{equation}
The $n=0$ moment, which corresponds to isospin charge, is not renormalized by
pion loops.
The parameter $\mu$ is introduced to suppress the rapid variation of
the logarithm for pion masses away from the chiral limit where
$\chi$PT breaks down.
Physically it is related to the size of the nucleon core, which acts as
the source of the pion field~\cite{Detmold:2001hq}.
Finally, the fits to the data are quite insensitive to the choice of $m_{b,n}$ 
(as long as it is large), and it has been set to 5~GeV for all $n$
\cite{Detmold:2001dv}.

A similar analysis leads to analogous lowest order LNA parameterizations
of the mass dependence of the spin-dependent moments \cite{CHINA}:
\begin{equation}
  \label{pxtrap}
  \langle x^n\rangle_{\Delta u-\Delta d}\ = \Delta a_n \left( 1 +
  \Delta c_{\rm LNA}m_\pi^2\log\frac{m_\pi^2}{m_\pi^2+\mu^2} \right)\
  +\ \Delta b_n\frac{m_\pi^2}{m_\pi^2+m_{b,n}^2}\,,
\end{equation}
and  
\begin{equation}  
  \label{txtrap}
  \langle x^n\rangle_{\delta u-\delta d}\ = \delta a_n \left( 1 +
  \delta c_{\rm LNA}m_\pi^2\log\frac{m_\pi^2}{m_\pi^2+\mu^2} \right)\
  +\ \delta b_n\frac{m_\pi^2}{m_\pi^2+m_{b,n}^2}\,,
\end{equation}
where the LNA coefficients are given by
$\Delta c_{\rm LNA}=-(1 + 2 g_A^2)/(4\pi f_\pi)^2$
and
$\delta c_{\rm LNA}=-(1 + 4 g_A^2)/[2(4\pi f_\pi)^2]$ \cite{CHPT}.
In the heavy quark limit, both $\Delta u (x)-\Delta d(x)$ and
$\delta u(x)-\delta d(x)$ are given by $\frac{5}{3}\delta(x-1/3)$
\cite{Gockeler:1996bm}, which leads to the constraints:
\begin{equation}
  \Delta b_n=\frac{5}{3^{n+1}} - \Delta a_n \left( 1 - \mu^2 \Delta
    c_{\rm LNA} \right)\,,
\end{equation}
and 
\begin{equation}
  \delta b_n=\frac{5}{3^{n+1}} - \delta a_n
  \left( 1 - \mu^2 \delta c_{\rm LNA} \right)\, .
\end{equation}
These are the most general lowest order parameterizations of the
twist-2 PDF moments consistent with chiral symmetry and the heavy
quark limits of QCD.

\subsection{Phenomenological Constraints}
\label{S:mu}

In Refs.~\cite{Detmold:2001jb,Detmold:2001dv} we presented analyses of
unpolarized data based on Eq.~(\ref{xtrap}), where it was concluded
that current lattice data alone do not sufficiently constrain the
extrapolation of these moments, and more accurate data at smaller
quark masses ($m_q\lesssim 15-20$~MeV) are required to determine the
parameter $\mu$.  In that work, a central value of $\mu=500$~MeV
(550~MeV when the heavy quark limit was not included
\cite{Detmold:2001jb}) was chosen as it best reproduced both the
lattice data and the phenomenological values at the physical point.
However, the systematic error on this parameter is very large; indeed,
the raw lattice data are consistent with $\mu=0$ (a linear
extrapolation).

In order to make the phenomenological constraint of $\mu$ more
quantitative, we employ the following measure of the goodness of fit
of the extrapolated values (at $m_\pi^{\rm phys}$) of the first three
non-trivial unpolarized moments to the phenomenological values,
$\langle x^i\rangle_{u-d}^{\rm expt}$, as a function of $\mu$:
\begin{equation}
    \label{E:goodfit}
    \chi(\mu)=\sum_{i=1}^3 \frac{(\langle x^i(\mu)\rangle_{u-d}- \langle
    x^i\rangle_{u-d}^{\rm expt})^2}{(\langle x^i\rangle_{u-d}^{\rm
    expt})^2}\, .
\end{equation}
\begin{figure}
  \begin{center}
  \includegraphics[width=14cm]{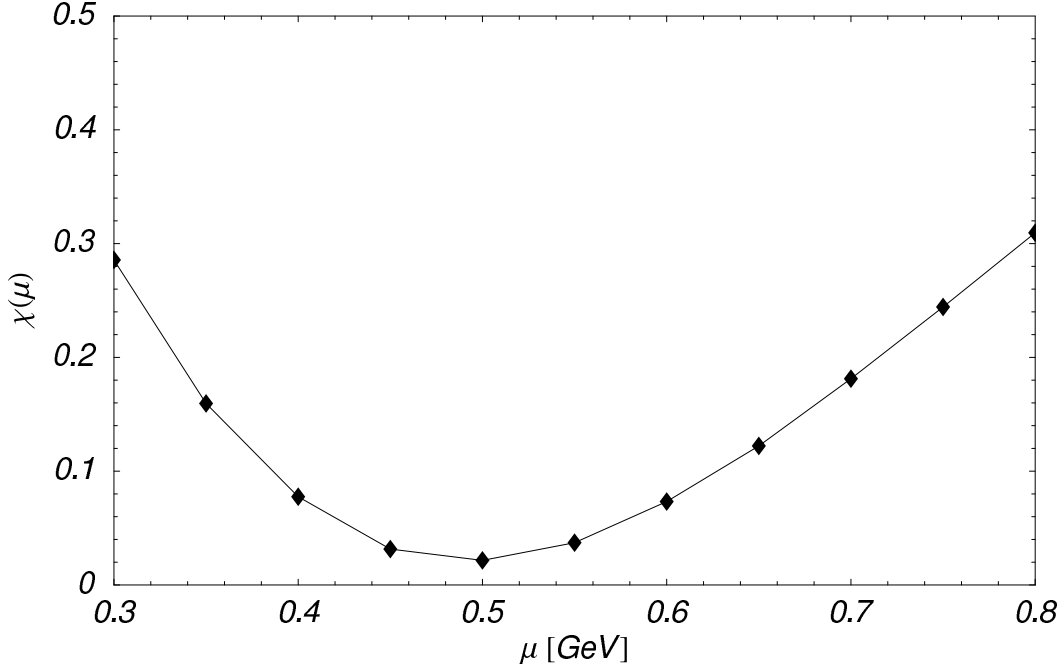}
    \caption{The goodness of fit of the extrapolated values of the
    first three non-trivial moments to the phenomenological values
    as a function of $\mu$ calculated using Eq.~(\ref{E:goodfit}).}
  \label{F:goodfit}
  \end{center}
\end{figure}
We assume that both the lattice data for the unpolarized moments and
their extrapolation based on Eq.~(\ref{xtrap}) are correct, and use the
phenomenological values for these moments as a constraint.

The behavior of the function $\chi(\mu)$ is shown in
Fig.~\ref{F:goodfit}, and the best value of $\mu$ is indeed found to
be 500~MeV.  This value is also comparable to the scale at which the
behavior found in other observables, such as magnetic moments and
masses, switches from smooth and constituent quark-like (slowly
varying with respect to the current quark mass) to rapidly varying and
dominated by Goldstone boson loops.  For fits to lattice data on
hadron masses, Leinweber {\em et al.} found values in the range 450 to
660~MeV \cite{MASS} when a sharp momentum cut-off was used.  The
similarity of these scales for the various observables is not
coincidental, but simply reflects the common scale at which the
Compton wavelength of the pion becomes comparable to the size of the
bare nucleon.  The value of $\mu$ is also similar to the scale
predicted by the $\chi^2$ fits to the model discussed in the following
section (see also Ref.~\cite{Detmold:2001jb}).

\subsection{$\Delta$ Intermediate States}
\label{S:deltas}

When we come to the calculation of polarized PDFs, there is considerable
phenomenological evidence to suggest that the $\Delta$ resonance will
play an important role.
Within the cloudy bag model (CBM), a convergent perturbative expansion of
the physical nucleon, in terms of the number of virtual pions, required
the explicit inclusion of the $\Delta$-isobar
\cite{Dodd:1981ve,Thomas:1982kv}
--- see also Ref.~\cite{Oettel:2002cw} for a recent, fully relativistic
investigation.
Without the $\Delta$, the ratio of the bare to renormalized pion-nucleon
coupling constant was found to be very large (as in the old Chew-Wick
model).
With it, they typically agree to within 10-15\%.
The essential physics is that the vertex renormalization associated with
coupling to the $\Delta$ or to an $N$--$\Delta$ transition compensates
almost exactly for the reduction caused by wave function renormalization.
Of course, the same mechanisms apply to the renormalization of the
axial charge, $g_A$, as to the pion nucleon coupling,
$g_{\pi NN}$.

In the limit that the $\Delta$ is degenerate with the nucleon,
$\Delta M \equiv M_\Delta - M \to 0$, the leading non-analytic
contribution from the $\Delta$ is of the same order as that arising from
the nucleon,
namely $m_\pi^2 \log m_\pi^2$.
In the limit $\Delta M \to \infty$, the $\Delta$ contribution can be
integrated out, and it formally does not make any non-analytic
contribution.
For a finite, but non-zero $\Delta M$, the vertex renormalization
involving the $\Delta$ is not a leading non-analytic term, but instead
enters as $m_\pi^3/\Delta M$.
However, the coefficient of this next-to-leading non-analytic (NLNA)
term is huge \cite{Chen:2001et} --- roughly three times bigger than
the $m_\pi^3$ term in the expansion of the nucleon mass \cite{MASS}.
{}Faced with such a large coefficient, one cannot rely on naive ordering
arguments alone to identify the important physics.

The solution adopted by Leinweber {\em et al.} \cite{MASS} in the
analysis of the chiral behavior of baryon masses was to calculate
corrections arising from those pion loop diagrams responsible for the
most rapid variation with $m_q$.  The finite spatial extension of the
pion source leads naturally to an ultraviolet cut-off at the $\pi N N$
and $\pi N \Delta$ vertices \cite{Hecht:2002ej}.  The parameter,
$\Lambda \sim 1/R$ (with $R$ the size of the source) associated with
these vertices is constrained phenomenologically.  This approach
ensures that the LNA and NLNA behavior of $\chi$PT is reproduced in
the $m_\pi \to 0$ limit, while the transition to the heavy quark limit
($m_\pi > \Lambda$), where pion loops are suppressed as inverse powers
of $m_\pi$, is also guaranteed.  Alternatively, one can study the
variation of PDF moments with $m_\pi$ within a model, such as the
cloudy bag \cite{Thomas:1982kv,Theberge:1981mq}, which also ensures
the full LNA and NLNA behavior of $\chi$PT, and in addition provides a
simple physical interpretation of the short-distance contributions (in
this case through the MIT bag model).  Rather than rely on a specific
model, in the present analysis we adopt the approach of
Ref.~\cite{MASS} and calculate the pion loop integrals with hadronic
vertices constrained phenomenologically.

\begin{figure}
  \begin{picture}(60,70)(20,0)
    \SetWidth{1.4}
    \Line(0,10)(100,10)
    \DashCArc(50,10)(35,0,180){5}
    \Text(50,-5)[cc]{$Z_2^N$}
  \end{picture}
\hspace{1.6cm}
  \begin{picture}(60,70)(20,0)
    \SetWidth{1.4}
    \Line(0,10)(100,10)
    \Line(15,12)(85,12)
    \DashCArc(50,10)(35,0,180){5}
    \Text(50,-5)[cc]{$Z_2^\Delta$}
  \end{picture}
\\ 
  \begin{picture}(60,70)(20,0)
    \SetWidth{1.4}
    \Line(0,10)(100,10)
    \DashCArc(50,10)(35,0,180){5}
    \CCirc(50,10){5}{Black}{White}
    \Line(46,6)(54,14)
    \Line(54,6)(46,14)
    \Text(50,-5)[cc]{$Z_{1,U/P}^{NN}$}
  \end{picture}
\hspace{1.6cm}
  \begin{picture}(60,70)(20,0)
    \SetWidth{1.4}
    \Line(0,10)(100,10)
    \Line(15,12)(85,12)
    \DashCArc(50,10)(35,0,180){5}
    \CCirc(50,11){5}{Black}{White}
    \Line(46,7)(54,15)
    \Line(54,7)(46,15)
    \Text(50,-5)[cc]{$Z_{1,U/P}^{\Delta\Delta}$}
  \end{picture}
\hspace{1.6cm}
  \begin{picture}(60,70)(20,0)
    \SetWidth{1.4}
    \Line(0,10)(100,10)
    \Line(50,12)(85,12)
    \DashCArc(50,10)(35,0,180){5}
    \CCirc(50,11){5}{Black}{White}
    \Line(46,7)(54,15)
    \Line(54,7)(46,15)
    \Text(50,-5)[cc]{$Z_{1,P}^{N\Delta}$}
  \end{picture}
\hspace{1.6cm}
  \begin{picture}(60,70)(20,0)
    \SetWidth{1.4}
    \Line(0,10)(100,10)
    \Line(15,12)(50,12)
    \DashCArc(50,10)(35,0,180){5}
    \CCirc(50,11){5}{Black}{White}
    \Line(46,7)(54,15)
    \Line(54,7)(46,15)
    \Text(50,-5)[cc]{$Z_{1,P}^{\Delta N}$}
  \end{picture}
\\ 
  \begin{picture}(60,80)(20,0)
    \SetWidth{1.4}
    \Line(0,10)(100,10)
    \DashCArc(50,10)(35,0,180){5}
    \CBox(81,6)(89,14){Black}{White}
    \Line(81,6)(89,14)
    \Line(89,6)(81,14)
    \Text(50,-5)[cc]{$Z_{1,P}^{N \rm WT}$}
  \end{picture}
\hspace{1.6cm}
  \begin{picture}(60,80)(20,0)
    \SetWidth{1.4}
    \Line(0,10)(100,10)
    \Line(15,12)(85,12)
    \DashCArc(50,10)(35,0,180){5}
    \CBox(81,6)(89,14){Black}{White}
    \Line(81,6)(89,14)
    \Line(89,6)(81,14)
    \Text(50,-5)[cc]{$Z_{1,P}^{\Delta \rm WT}$}
  \end{picture}
\hspace{1.6cm}
  \begin{picture}(60,80)(20,0)
    \SetWidth{1.4}
    \Line(0,10)(100,10)
    \DashCArc(50,30)(20,0,360){5}
    \CCirc(50,10){5}{Black}{White}
    \Line(46,6)(54,14)
    \Line(54,6)(46,14)
    \Text(50,-5)[cc]{$Z_{1,U/P}^{\rm tad}$}
  \end{picture}
\\
\caption{\label{FZcontribs} Contributions to the wave function and
  vertex renormalization of the nucleon matrix elements of the operators
  ${\cal O}_i^{\mu_1\ldots\mu_n}$, $i=q, \Delta q, \delta q$,
  in Eq.~(\protect{\ref{E:operators}}).
  Solid, double and dashed lines denote nucleon, $\Delta$ and pion
  propagators and the crossed circle and box indicate the insertion of the
  relevant operators. Diagrams $Z_2^N$ and $Z_2^\Delta$ denote the
  contributions to wave function renormalization (a derivative with
  respect to the external momentum is implied).}
\end{figure}
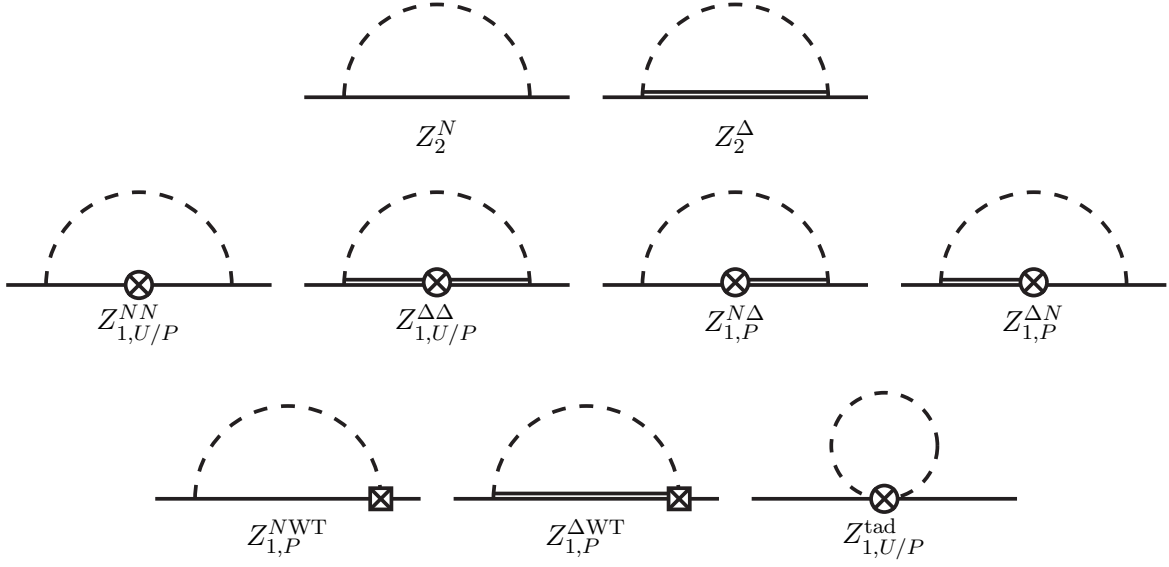

The overall renormalization of the forward matrix elements of the
operators of Eqs.~(\ref{E:operators}) in nucleon states is then given
by:
\begin{equation}
\label{merenorm}
  \langle N|{\cal O}^{\mu_1\ldots\mu_n}_i
  |N\rangle_{\rm dressed}
= \frac{Z_2}{Z_i}
  \langle N|{\cal O}^{\mu_1\ldots\mu_n}_i
  |N\rangle_{\rm bare}\ , \ \ \ \ i=q,\Delta q,\delta q\ ,
\end{equation}
where $Z_2$ is the wave function renormalization constant,
\begin{equation}
  Z_2^{-1}= 1 + Z_2^N + Z_2^\Delta\, ,
\end{equation}
and $Z_i$ are the vertex renormalization constants described below.
The $N$ and $\Delta$ contributions to the wave function
renormalization, illustrated in the first row of
Fig.~\ref{FZcontribs}, are given in the heavy baryon
limit\footnote{While the heavy baryon limit applies strictly when
  $m_\pi << M$, the form factor, $u(k)$, strongly suppresses all of
  these integrals for $m_\pi$ above 400 MeV and thus the heavy baryon
  expression provides an adequate description of the meson loops in
  the region where they are large and rapidly varying.} by
\cite{Theberge:1981mq}:
\begin{eqnarray}
  Z_2^{N} &=& \frac{3g_A^2}{(4\pi f_\pi)^2}\int_0^\infty\frac{k^4
      u^2(k)dk}{\omega^3(k)}\,,
\\
  Z_2^{\Delta} &=&
  \frac{4}{9}\frac{g_{\pi N\Delta}^2}{g_{\pi NN}^2}
  \frac{3g_A^2}{(4\pi f_\pi)^2}\int_0^\infty\frac{k^4
      u^2(k)dk}{\omega(k)(\omega(k)+\Delta M)^2}\ ,
\end{eqnarray}
where $\omega(k)=\sqrt{k^2+m_\pi^2}$ is the pion energy, and $u(k)$ is
the form factor parameterizing the momentum dependence of the $\pi NN$
and $\pi N \Delta$ vertices, for which we choose a dipole form,
\begin{eqnarray}
u(k) &=& \frac{\Lambda^4}{(k^2+\Lambda^2)^2}\ .
\end{eqnarray}
The numerical calculations are performed with a characteristic momentum
cut-off scale $\Lambda=0.8$~GeV, just a little softer than the measured
axial form factor of the nucleon \cite{Thomas:tv,Guichon:1982zk}
-- although the results are relatively
insensitive to the precise value of $\Lambda$, as illustrated below.
The ratio of the $\pi N\Delta$ to $\pi NN$ couplings can be determined
from SU(6) symmetry ($g_{\pi N\Delta}/g_{\pi NN}=\sqrt{72/25}$), however,
in the numerical calculations we consider a range of values for the
ratio.
SU(6) symmetry is also used to relate matrix elements of the twist-2
operators in the bare $\Delta$ and $N$-$\Delta$ transition to those in
the bare nucleon.
Lattice calculations of $\Delta$ or $N$-$\Delta$ transition matrix elements
will in future test the reliability of this approximation.

The renormalization constants for the spin-independent, helicity and
transversity operators are given by
\begin{subequations}
\begin{eqnarray}
  Z_{q}^{-1}
  &=& 1+Z_{1,U}^{NN}+ Z_{1,U}^{\Delta\Delta} + Z_{1,U}^{\rm tad}\, , \\
  Z_{\Delta q}^{-1}
  &=& 1 + Z_{1,P}^{NN} + Z_{1,P}^{N\Delta} + Z_{1,P}^{\Delta N}
   + Z_{1,P}^{\Delta\Delta} + Z_{1,P}^{\rm tad} + Z_{1,P}^{N \rm WT}
   + Z_{1,P}^{\Delta \rm WT}\, ,                                \\
  Z_{\delta q}^{-1}
  &=& 1 + Z_{1,P}^{NN} + Z_{1,P}^{N\Delta} + Z_{1,P}^{\Delta N}
   + Z_{1,P}^{\Delta\Delta} + \frac{1}{2} Z_{1,P}^{\rm tad}
   + \frac{1}{2} Z_{1,P}^{N \rm WT}
   + \frac{1}{2} Z_{1,P}^{\Delta \rm WT}\, .
\end{eqnarray}
\end{subequations}%
The contributions from the coupling to nucleon intermediate states are
given by:
\begin{eqnarray}
  Z_{1,U}^{NN} &=& -\frac{g_A^2}{(4\pi f_\pi)^2}
  \int_0^\infty\frac{k^4 u^2(k)dk}{\omega^3(k)}\,,
\end{eqnarray}
and
\begin{eqnarray}
  Z_{1,P}^{NN} &=& \frac{1}{3}\frac{g_A^2}{(4\pi f_\pi)^2}
  \int_0^\infty\frac{k^4 u^2(k)dk}{\omega^3(k)}\,,
\end{eqnarray}
for the unpolarized and polarized operators, respectively.
One can explicitly verify that the LNA behavior of these contributions
is $m_\pi^2 \log m_\pi^2$.
The $\Delta$ contributions to the unpolarized and polarized operators
are equivalent,
\begin{eqnarray}
  Z_{1,U}^{\Delta\Delta} = Z_{1,P}^{\Delta\Delta} &=&
  \frac{20}{27}\frac{g_{\pi N\Delta}^2}{g_{\pi NN}^2} 
  \frac{g_A^2}{(4\pi f_\pi)^2}\int_0^\infty\frac{k^4
    u^2(k)dk}{\omega(k)(\omega(k)+\Delta M)^2}\ ,
\end{eqnarray}
while the $N\Delta$ transition contributes only to the spin-dependent
operators,
\begin{eqnarray}
  Z_{1,P}^{\Delta N} = Z_{1,P}^{N\Delta} &=&
  \frac{16}{27}\frac{g_{\pi N\Delta}^2}{g_{\pi NN}^2}
  \frac{g_A^2}{(4\pi f_\pi)^2}\int_0^\infty\frac{k^4 
    u^2(k)dk}{\omega^2(k)(\omega(k)+\Delta M)}\ .
\end{eqnarray}
These contributions are illustrated in the middle row in
Fig.~\ref{FZcontribs}.
Expanding these terms for small $m_\pi$, one finds that the leading
non-analytic terms associated with the $\Delta$ and $N$--$\Delta$
transition contributions enter at orders $m_\pi^4 \log m_\pi^2$ and
$m_\pi^3$, respectively.
The contributions from the tadpole diagrams, which are independent of
$g_A$, are also identical for the unpolarized and polarized cases, and
given by
\begin{eqnarray}
\label{tadpole}
  Z_{1,U}^{\rm tad}\ =\ Z_{1,P}^{\rm tad}
  &=& -\frac{2}{(4\pi f_\pi)^2}
    \int_0^\infty\frac{k^3 u^2(k)dk}{\omega^2(k)}\ .
\end{eqnarray}
The tadpole contributions also enter at order $m_\pi^2 \log m_\pi^2$
\cite{CHPT}, as can be verified directly from Eq.~(\ref{tadpole}).

While the inclusion of the $\Delta$ resonance is important for
quantitative descriptions of baryon structure, we also know from
phenomenological studies that the higher order (in $m_\pi$)
Weinberg-Tomozawa contact term \cite{Weinberg:1966fm,Tomozawa:1966gg}
plays a vital role in low energy $S$-wave pion--nucleon scattering
\cite{Thomas:1981ps}.
Because of the Adler-Weisberger relation \cite{AW} between $\pi N$ cross
sections and $g_A$, any term which affects $\pi N$ cross sections should
also have some effect on $g_A$ \cite{Morgan:1985kr}.
In fact, within the CBM Morgan {\em et al.} \cite{Morgan:1985kr} found
that this term largely resolves the discrepancy between the bag model
value of $g_A=1.09$ and the empirical value of $g_A$ for bag radii
$R \in (0.9,1.1)$~fm.
In the present treatment, since we do not use the CBM explicitly, but
rather parameterize the pion source via the phenomenological
form factor $u(k)$, we determine the overall strength of the
Weinberg-Tomozawa term so as to reproduce the contribution found in the
CBM, as outlined in Ref.~\cite{Morgan:1985kr}.
The relative contributions of the diagrams with $N$ and $\Delta$
intermediate states, however, illustrated in the last row of
Fig.~\ref{FZcontribs}, can be fixed by SU(6) symmetry.
These contributions to the operator renormalization can then be written:
\begin{eqnarray}
  Z_{1,P}^{N \rm WT} &=& C_{WT}\int_0^\infty\frac{k^4
    u^2(k)dk}{\omega^2(k)}\ ,
\label{eq:WT1}
\\
  Z_{1,P}^{\Delta \rm WT} &=& \frac{C_{WT}}{18}(1+\sqrt{2})\frac{g_{\pi
      N\Delta}^2}{g_{\pi NN}^2}\int_0^\infty\frac{k^4 
    u^2(k)dk}{\omega(k)(\omega(k)+\Delta M)}\ ,
\label{eq:WT}
\end{eqnarray}
for the $N$ and $\Delta$ intermediate states, respectively, where
$C_{WT}$ is the overall normalization.
{}For the above range of $R$, the physical value of $g_A$ can be
reproduced to within a few percent for the corresponding range of
$C_{WT} \in (0.21,0.30)$.
In the following numerical analysis, we use this range as an estimate of
the systematic error on the Weinberg-Tomozawa contribution.
Even though the non-analytic behavior of the integrals in
Eqs.~(\ref{eq:WT1}) and (\ref{eq:WT})
is $m_\pi^3$ or higher, their contributions are found to be significant.
Note, however, that the Weinberg-Tomozawa terms contribute only to
spin-dependent matrix elements, and make no contribution to unpolarized
matrix elements.

With the exception of the matrix elements of the unpolarized, $n=0$
operator, the renormalization of each moment of the various
distributions is independent of $n$.
The $n=0$ operator, which corresponds to the isospin charge, is not
renormalized --- additional contributions from operator insertions on
the pion propagator cancel those shown in Fig.~\ref{FZcontribs}.

\begin{figure}
  \includegraphics[width=13.5cm]{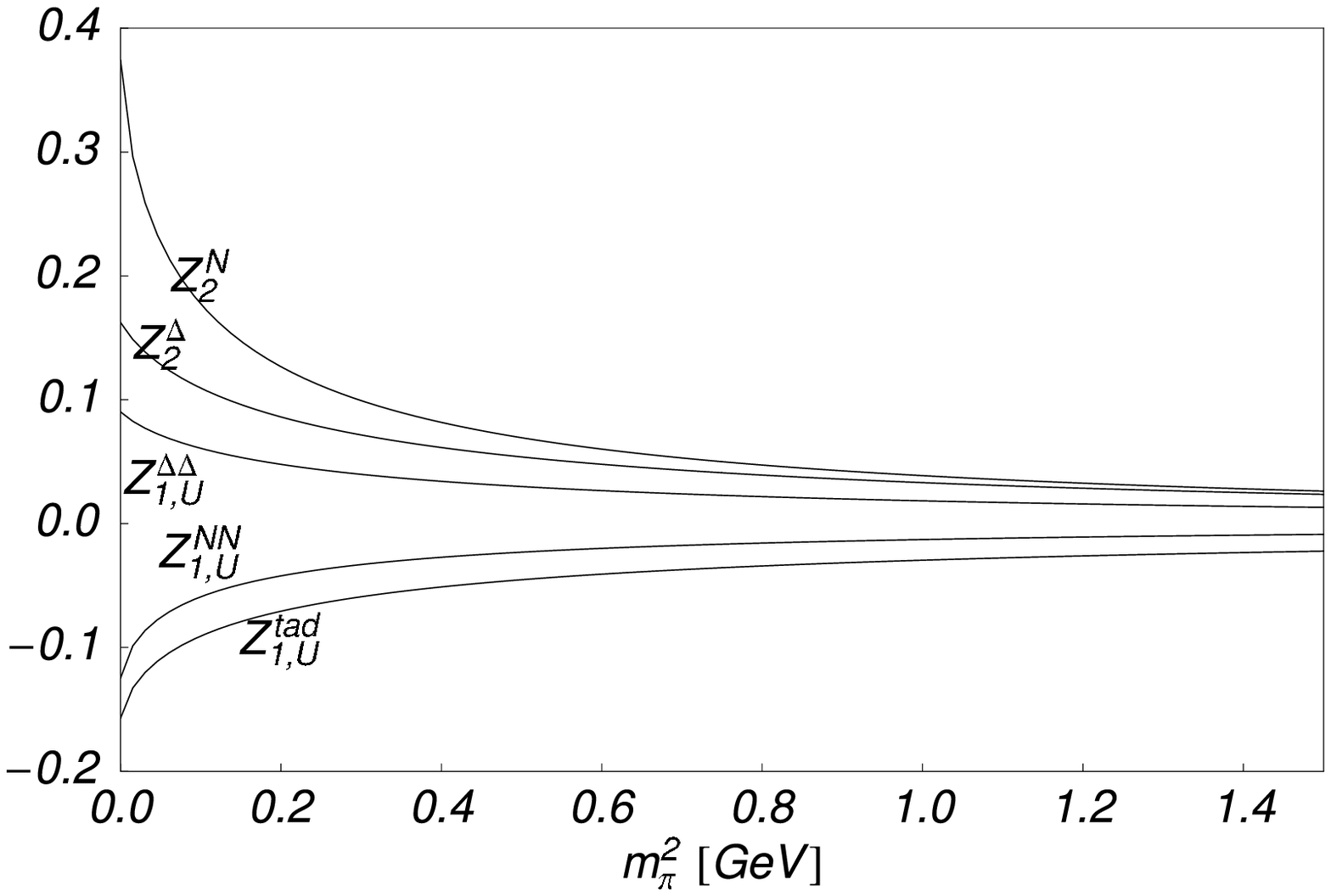}
  \includegraphics[width=13.5cm]{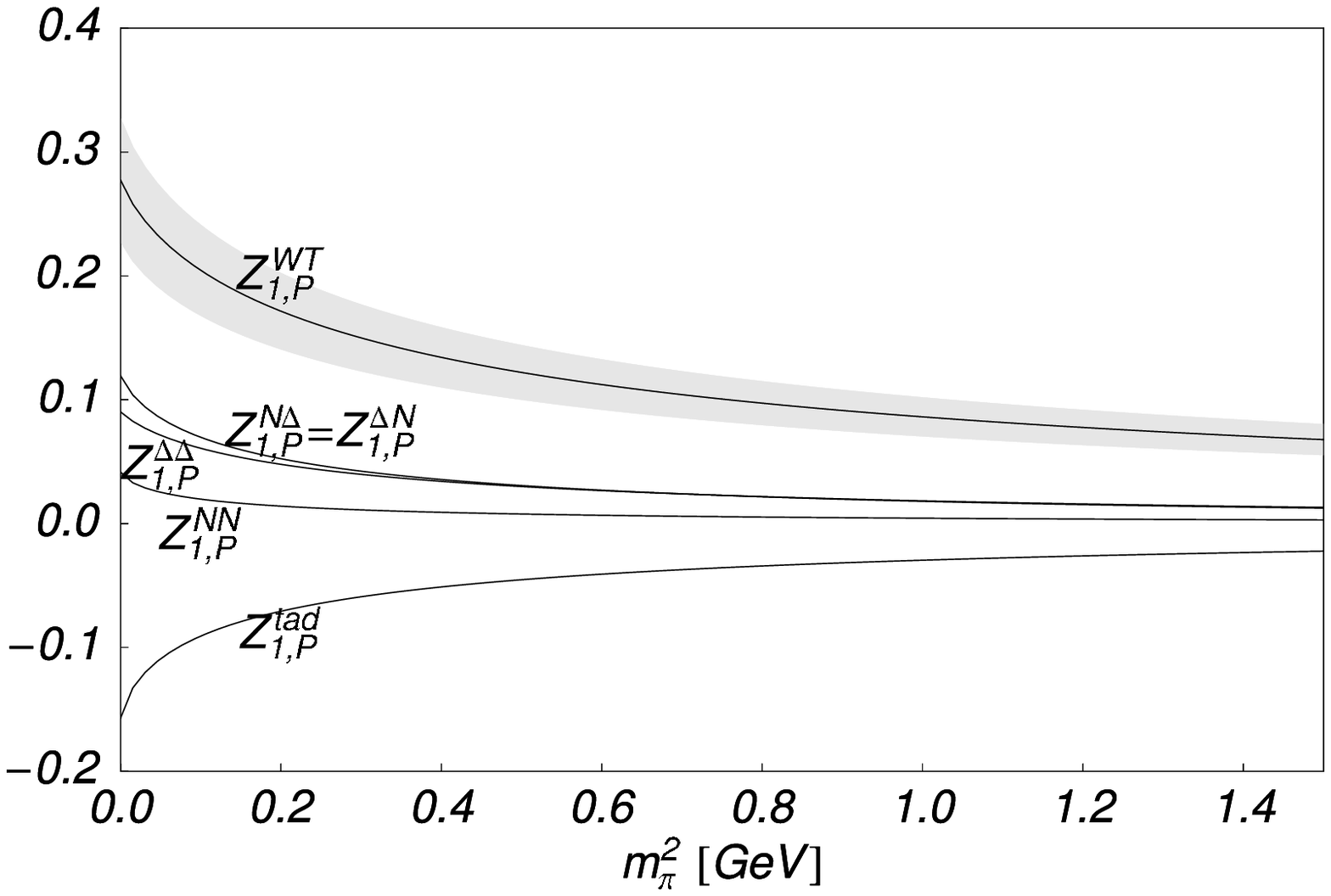}
  \caption{\label{Zcontribs}
    Contributions to the pion loop renormalization of the matrix
    elements of the twist--2 operators required to evaluate the
    moments of the PDFs.  The upper panel shows nucleon wave function
    renormalizations ($Z_2^N$, $Z_2^\Delta$) and spin-independent
    operator renormalizations.  The lower panel shows the
    contributions to the renormalization of spin-dependent operators,
    and the shaded region is an estimate of the uncertainty in the
    Weinberg--Tomozawa term, $Z^{\rm WT}_{1,P} \equiv Z^{\rm N
      WT}_{1,P} + Z^{\rm \Delta WT}_{1,P}$.  The $g_{\pi
      N\Delta}/g_{\pi NN}$ coupling constant ratio is set to the SU(6)
    symmetric value of $\sqrt{72/25}$.}
\end{figure}

The pion mass dependence of the various contributions to the wave
function and operator renormalization is shown in Fig.~\ref{Zcontribs}.
For the ratio of the couplings, $g_{\pi N\Delta}/g_{\pi NN}$, SU(6)
symmetry is assumed.
The relative size of the terms $Z_1^{NN}$ and $Z_1^{\Delta\Delta}$ in the
spin-dependent and spin-independent cases already makes it clear that
intermediate states involving $\Delta$ resonances are much more
significant in the former case.
In particular, whereas $Z_{1,P}^{NN}$ does little to counter the effect
of the wave function renormalization, the $\Delta$ contributions
$Z_{1,P}^{\Delta\Delta}$, $Z_{1,P}^{N\Delta}$, and $Z_{1,P}^{\Delta N}$
essentially cancel its effect.
\begin{figure}
  \includegraphics[width=14cm]{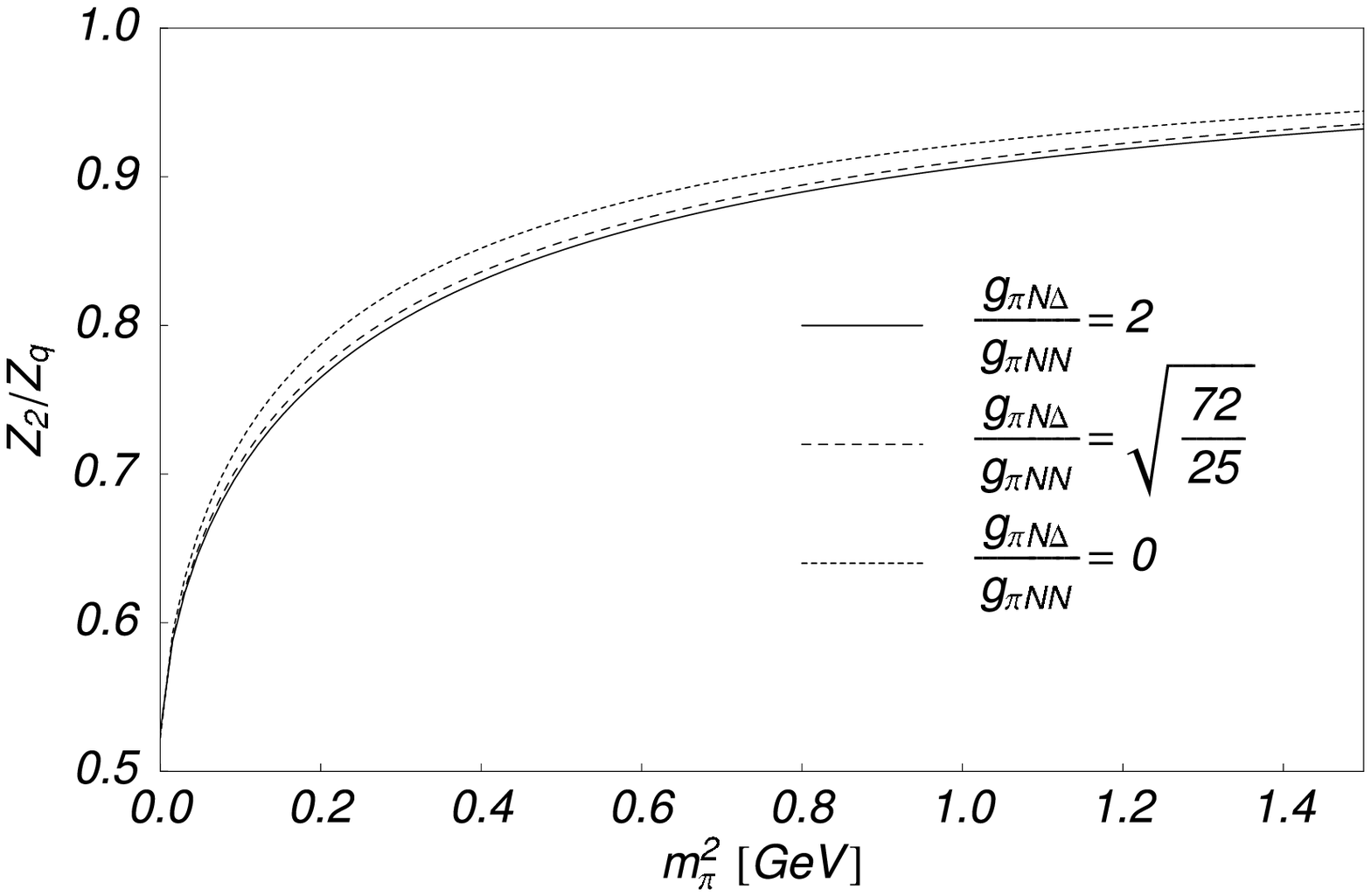}
  \includegraphics[width=14cm]{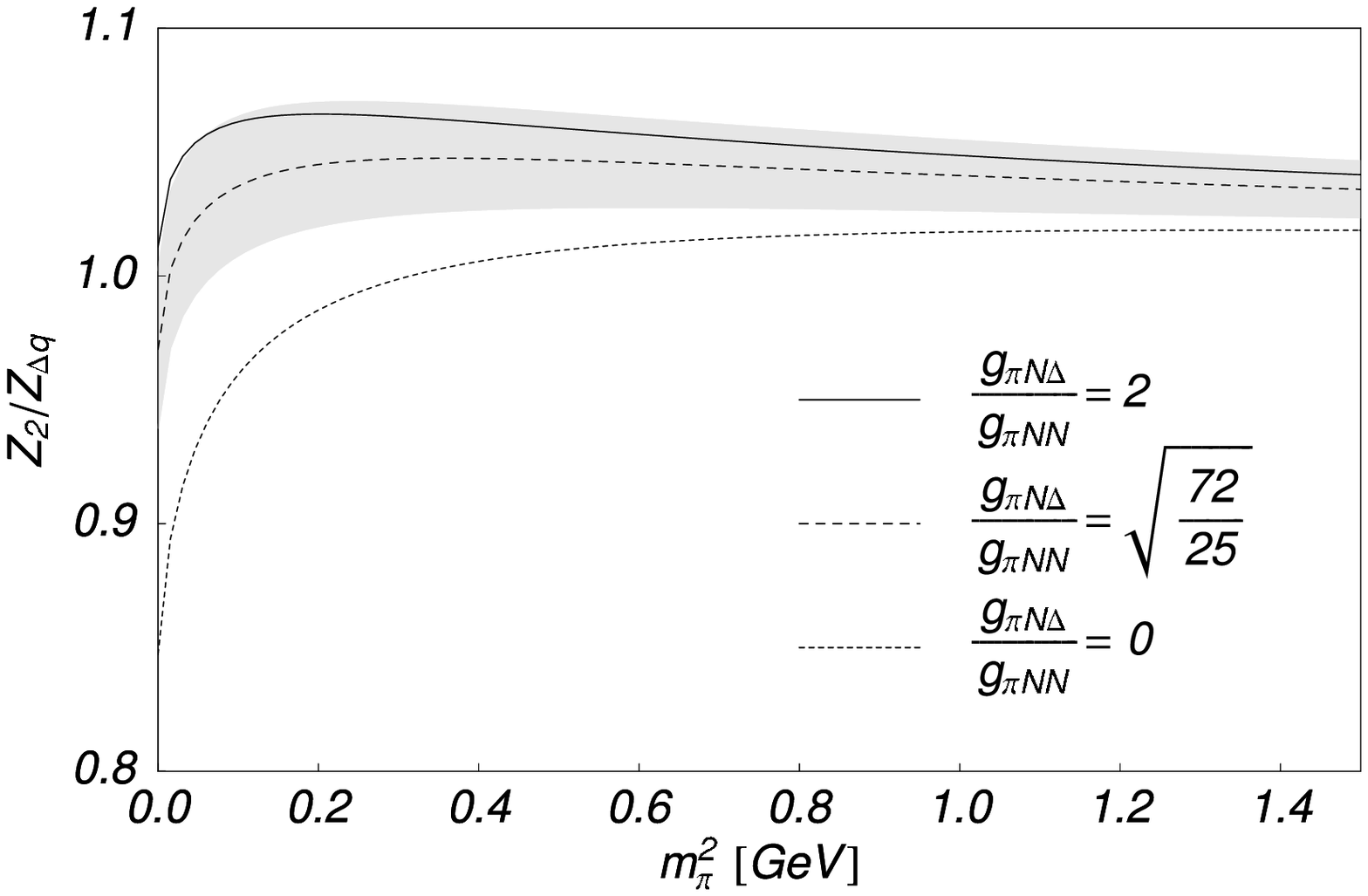}
  \caption{\label{Z2overZ1}
        Pion dressing of the matrix elements of the spin-independent
        (upper panel) and spin-dependent (lower panel) operators in
        Eq.~(\protect{\ref{E:operators}}) for various values of the ratio
        of coupling constants, $g_{\pi N\Delta}/g_{\pi NN}$.
        The shading in the lower panel indicates the variation about the
        dashed curve ($g_{\pi N\Delta}/g_{\pi NN}=\sqrt{72/25}$) caused by
        the uncertainty in the Weinberg--Tomozawa term. 
        The behavior of $Z_2/Z_{\delta q}$ is similar to that of
        $Z_2/Z_{\Delta q}$.}
\end{figure}

\begin{figure}
\includegraphics[width=14cm]{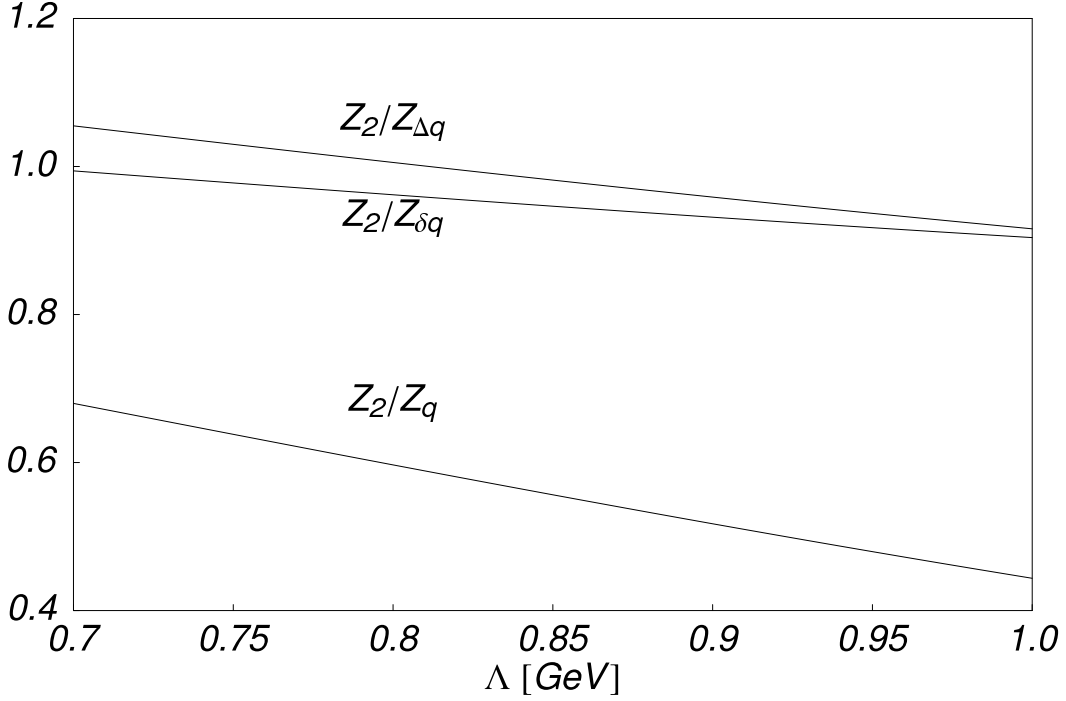}
\caption{Dependence of the renormalization of the nucleon matrix elements
        in Eq.~(\ref{merenorm}) on the dipole mass parameter, $\Lambda$,
        at the physical pion mass, for the SU(6) value of
        $g_{\pi N\Delta}/g_{\pi NN}$.}
\label{fig:Z2Z1vsLambda}
\end{figure}

To explore the sensitivity of the results to the strength of the
$\Delta$ contribution, in Fig.~\ref{Z2overZ1} we show the combined
effect of the pion dressing on spin-averaged (upper panel) and
spin-dependent (lower panel) nucleon matrix elements in
Eq.~(\ref{merenorm}) for a range of values of the ratio $g_{\pi
  N\Delta}/g_{\pi NN}$.  For illustration we choose values of $g_{\pi
  N\Delta}/g_{\pi NN}$ equal to zero (no $\Delta$ states),
$\sqrt{72/25}$ (SU(6) coupling) and 2 (phenomenological value needed
to reproduce the width of the physical $\Delta$ resonance).  In the
unpolarized case, the effect of this variation is relatively small ---
less than 3\% over the entire range of pion masses considered here.
In contrast, the effect of the $\Delta$ on the helicity and
transversity moments (matrix elements of the spin-dependent operators
${\cal O}_{\Delta q}^{\mu_1\ldots\mu_n}$ and ${\cal O}_{\delta
  q}^{\mu_1\ldots\mu_n}$) is far more significant.  If the
contribution from the $\Delta$ (and the $N$-$\Delta$ transition matrix
elements) is ignored ($g_{\pi N\Delta}/g_{\pi NN}=0$), $\left.
  Z_2/Z_{\Delta q}\right|_{m_\pi^{\rm phys}}=0.90$, while including
these contributions with SU(6) couplings increases this to $\left.
  Z_2/Z_{\Delta q}\right|_{m_\pi^{\rm phys}}=1.01$, and to $\left.
  Z_2/Z_{\Delta q}\right|_{m_\pi^{\rm phys}}=1.04$ at the
phenomenological value $g_{\pi N\Delta}/g_{\pi NN}=2$.  Consequently,
when the effects of the $\Delta$ are included with a coupling constant
which is consistent with phenomenology, one finds that there is almost
no curvature in the extrapolation of the spin-dependent moments.
This result is relatively stable against variations
\cite{Thomas:tv,Guichon:1982zk}
in the dipole mass parameter, $\Lambda$, in the range $\sim 0.7-1.0$~GeV
--- especially for the spin-dependent moments, as illustrated in
Fig.~\ref{fig:Z2Z1vsLambda}.

Matrix elements of the twist-2 operators (\ref{E:operators}) in
\emph{bare} nucleon states will necessarily be analytic functions of
the quark mass ($m_q \sim m_\pi^2$), so the one pion loop
renormalization described above is the only possible source of LNA
contributions.
Consequently, the LNA behavior of the matrix elements in
Eq.~(\ref{merenorm}) will be given by
\begin{equation} 
\langle {\cal O}^{\mu_1\ldots\mu_n}_i\rangle_{\rm LNA}
 = Z_2^{\rm LNA} - Z_i^{\rm LNA}\ ,\ \ \ \ \ i=q,\Delta q,\delta q\ .
\end{equation}
If the $N$--$\Delta$ mass splitting is artificially reduced to zero,
$\Delta$ intermediate states become degenerate with the corresponding
nucleon intermediate states, and the respective $\Delta$ diagrams
formally give rise to LNA contributions.
Leaving the $g_{\pi N\Delta}/g_{\pi NN}$ ratio free, the coefficients of
the LNA contributions (the $m_\pi^2 \log m_\pi^2$ term) to the various
matrix elements can then be written:
\begin{subequations}
\begin{eqnarray}
  \langle {\cal O}^{\mu_1\ldots\mu_n}_{q}\rangle_{\rm LNA}  
  = Z_{2}^{\rm LNA} - Z_{q}^{\rm LNA}
  = -\frac{1}{(4\pi f_\pi)^2}\left[\left(3
      +\frac{16}{27}\frac{g^2_{\pi N\Delta}}{g^2_{\pi
          NN}}\right)g^2_{A}+1\right]\, ,           \\
  \label{LNABehave_hel}
  \langle {\cal O}^{\mu_1\ldots\mu_n}_{\Delta q}\rangle_{\rm
    LNA}  
  = Z_{2}^{\rm LNA} - Z_{\Delta q}^{\rm LNA}
  = -\frac{1}{(4\pi f_\pi)^2}\left[\left(2 
      -\frac{4}{9}\frac{g^2_{\pi N\Delta}}{g^2_{\pi
          NN}}\right)g^2_{A}+1\right]\, ,           \\
  \label{LNABehave_trans}
  \langle {\cal O}^{\mu_1\ldots\mu_n}_{\delta q}\rangle_{\rm
    LNA}  
  = Z_{2}^{\rm LNA} - Z_{\delta q}^{\rm LNA}
  = -\frac{1}{(4\pi f_\pi)^2}\left[\left(2 
      -\frac{4}{9}\frac{g^2_{\pi N\Delta}}{g^2_{\pi
          NN}}\right)g^2_{A}+\frac{1}{2}\right]\, .
\end{eqnarray}
\end{subequations}
This makes it clear that, whereas an increase in
$g_{\pi N\Delta}/g_{\pi NN}$ from $0$ (no $\Delta$ contributions) tends
to increase the effective coefficient of the chiral logarithm in the
unpolarized case, for the spin-dependent operators it acts to suppress it.
Indeed, assuming the bare axial coupling $g_{A}=1.26$,
at $g_{\pi N\Delta}/g_{\pi NN}=2.43$ the LNA coefficient for the
polarized moments is {\it zero}, and for larger values it even becomes 
{\it positive}.
Whilst this exact cancellation is an artifact of setting $\Delta M=0$,   
it highlights the significant role played by the $\Delta$ resonance.

{}From this analysis and the numerical results shown earlier, one can
conclude that the inclusion of the $\Delta$ resonance will cause only a
minor quantitative change in the extrapolation of unpolarized moments,
and in practical extrapolations of lattice data the $\Delta$ can be
neglected with no loss of accuracy, given the current uncertainties
in the data.
In contrast, the $\Delta$ leads to a qualitatively different picture
for the extrapolation of the spin-dependent moments and {\it must} be
included there.

There are a number of possible approaches that can be taken to account
for the $\Delta$ contributions.
One strategy would be to include the one-loop renormalizations
numerically in the extrapolations, along the lines of the calculation
of self-energies in the hadron mass extrapolations in
Refs.~\cite{MASS,Young:2001nc}.
One could also replace the momentum integrals in the expressions for
$Z_1$ and $Z_2$ with discrete sums over momenta which are available on
the lattice, $\int d^3k \to (1/V) (2\pi/a)^3 \sum_{k_x,k_y,k_z}$,
where $V$ is the spatial volume of the lattice, as in the analysis of
the $\rho$ meson mass in Ref.~\cite{DISCRETE} (see also
\cite{QUENCH}).  Because of the discretization of space-time on the
lattice, the lattice momenta are restricted to values $k_\mu = 2\pi
n_\mu/a L_\mu$, where $L_\mu$ is the number of lattice sites in the
$\mu$ direction and the integer $n_\mu$ runs between $-L_\mu/2$ and
$+L_\mu/2$.  We have checked that at large $m_\pi^2$ the differences
between the integral and discrete sum are only a few percent or less,
however, at small $m_\pi^2$ values the momentum gap between $k_\mu=0$
and the minimum momentum allowed, $k_\mu=\pm 2\pi/a L_\mu$, may
introduce corrections.

\begin{figure}
  \begin{center} \includegraphics[width=14cm]{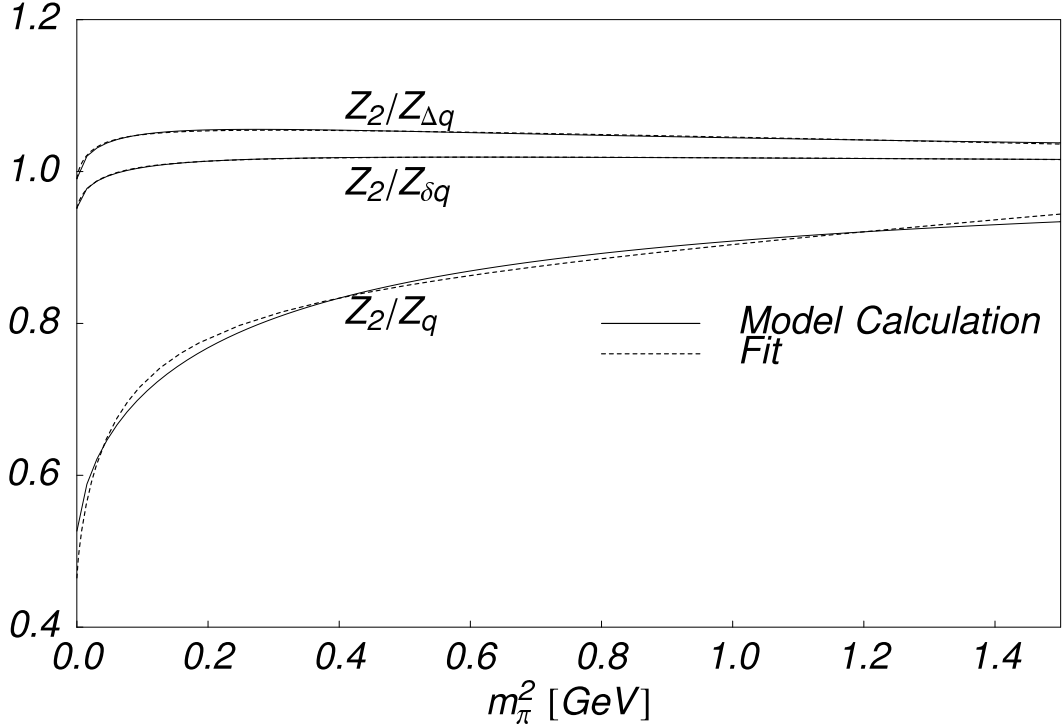}
    \caption{\label{Ffit} Fits to the calculated renormalization factors
    $Z_2/Z_i,\ i=q,\Delta q,\delta q$, in Eq.~(\protect\ref{merenorm})
    using the functional form in Eq.~(\protect\ref{renfit}), as a
    function of $m_\pi^2$.  The $\pi N\Delta$ to $\pi NN$ coupling ratio
    has been set to the average of the SU(6) and phenomenological values
    ($g_{\pi N\Delta}/g_{\pi NN}=1.85$), and $g_{A}$ to the tree
    level value, 1.26.  The corresponding values of $\mu$ are given in
    the third row of Table~\protect\ref{fittab}.}
  \end{center}
\end{figure}

Although this procedure is more accurate in principle, in practical
extrapolations of lattice data it is not as straightforward to
implement as an extrapolation formula based on a simple functional
form would be.  For this purpose it is more useful to preserve the
simplicity of a single formula which interpolates between the distinct
realms of chiral perturbation theory and contemporary lattice
simulations, as proposed in
Refs.~\cite{Detmold:2001jb,Detmold:2001dv}.
In order to test whether one can continue to apply a modified form of
the extrapolation formula in Eq.~(\ref{xtrap}) to lattice data for the
spin-dependent moments, as well as the spin-independent,
we attempt to fit the pion mass dependence of the
renormalization factors shown in Fig.~\ref{Z2overZ1} using the form
\begin{equation}
  \label{renfit}
  Z_2/Z_i = \alpha_i + \beta_i m_\pi^2 + 
  \frac{\gamma_{i, \rm LNA}}{(4\pi f_\pi)^2} m_\pi^2
  \log\left[\frac{m_\pi^2}{m_\pi^2+\mu_i^2}\right]\ ,\ \ \ \ \
i=q,\Delta q,\delta q\ ,
\end{equation}
with $\alpha_i$, $\beta_i$ and $\mu_i$ treated as free parameters, but
with $\gamma_{i, \rm LNA}$ fixed to the values obtained analytically
in the limit $\Delta M \to 0$, as shown in Table~\ref{fittab}.  The
fits to $Z_2/Z_i$ $(i=q,\Delta q,\delta q)$ are illustrated in
Fig.~\ref{Ffit} for the average of the $g_{\pi N\Delta}/g_{\pi NN}$
values from SU(6) symmetry ($\sqrt{72/25}$) and phenomenology (2).
%
%
Fits for other values of the coupling are equally good. It is
remarkable that the LNA form (\ref{renfit}) is indeed able to
reproduce the full calculations of $Z_2/Z_i$ with such high accuracy,
given that the full calculations include higher order effects (in
$m_\pi$) associated with the $\Delta$ and Weinberg-Tomozawa
contributions.  The best fit values of $\mu$, shown in
Table~\ref{fittab}, are only slightly smaller than those found in
earlier work \cite{Detmold:2001jb,Detmold:2001dv}.
Note that the functional form in Eq.~(\ref{renfit}) does not include
the modifications designed to ensure the correct heavy quark limit,
as in Eqs.~(\ref{xtrap})--(\ref{txtrap}) --- incorporating this
constraint leads to only marginal changes in the parameter $\mu$
\cite{Detmold:2001dv}.
\begin{table}
  \begin{ruledtabular}
    \begin{tabular}{ccccccc}
      &\multicolumn{2}{c}{$Z_2/Z_q$} &\multicolumn{2}{c}{$Z_2/Z_{\Delta
          q}$} &\multicolumn{2}{c}{$Z_2/Z_{\delta q}$} \\
      $g_{\pi N\Delta}/g_{\pi NN}$ & $c_{\rm LNA}$ & $\mu$ (GeV)
        & $\Delta c_{\rm LNA}$ & $\mu$ (GeV) & $\delta c_{\rm LNA}$
        & $\mu$ (GeV) \\ 
      \hline
      0 & $3g_{A}^2+1$ & 0.45 & $2g_{A}^2+1$ & 0.28 &
      $2g_{A}^2+\frac{1}{2}$ & 0.32 \\
      $\sqrt{\frac{72}{25}}$ & $\frac{107}{25}g_{A}^2+1$ & 0.39 &
      $\frac{18}{25}g_{A}^2+1$ & 0.25 &
      $\frac{18}{25}g_{A}^2+\frac{1}{2}$ & 0.29 \\ 
      1.85 & $4.51g_{A}^2+1$ & 0.38 &
      $0.48 g_{A}^2+1$ & 0.25 &
      $0.48 g_{A}^2+\frac{1}{2}$ & 0.30 \\ 
      2 & $\frac{43}{9}g_{A}^2+1$ & 0.37 &
      $\frac{2}{9}g_{A}^2+1$ & 0.24 &
      $\frac{2}{9}g_{A}^2+\frac{1}{2}$ & 0.29 \\ 
    \end{tabular}
  \end{ruledtabular}
  \caption{\label{fittab}
        Fits to the dependence on $m_\pi^2$ of the calculated
        renormalization factors, obtained by varying $\alpha_i$, $\beta_i$
        and $\mu_i$ in Eq.~(\protect\ref{renfit}).  The LNA coefficients
        and the mass parameters $\mu_i$ are shown for various values of
        the $g_{\pi N\Delta}/g_{\pi NN}$ ratio: 0 (no $\Delta$),
        $\sqrt{72/25}$ (SU(6)), 2 (phenomenological value) and
        1.85 (average of SU(6) and phenomenological values).}
\end{table}

As discussed above, excluding the isospin charge, all moments of each
operator are renormalized in the same manner.
Hence, our conclusions regarding the inclusion of the $\Delta$-isobar
apply equally well to extrapolations of
$g_A = \langle 1\rangle_{\Delta u-\Delta d}$ and all other moments of the
helicity and transversity distributions.

\section{Extrapolation of Lattice Data}
\label{S:xtrap}

\begin{figure}
\begin{center}
\includegraphics[width=13.cm]{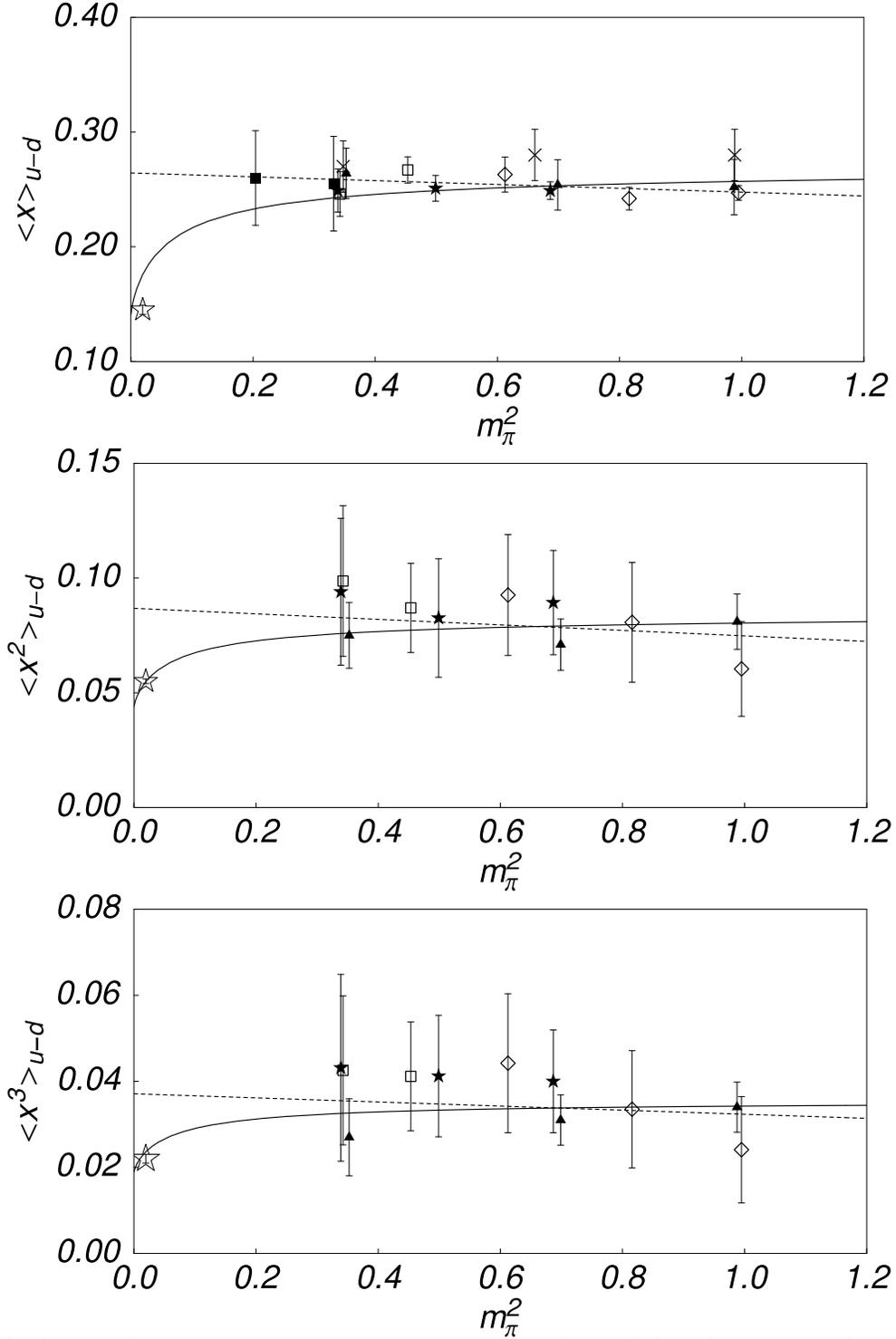}
\vspace{-0.8cm}
\caption{The lowest three non-trivial moments of the unpolarized
  distribution $u - d$, extrapolated using a naive linear fit
  (dashed lines) and the improved chiral extrapolation (solid lines).
  The stars indicate the experimentally measured moments at the physical
  pion mass, and the lattice data are taken from the sources listed in
  Table~\protect\ref{T:data}, where the various plotting symbols are
  defined.}
\label{F:unpolarized}
\end{center}
\end{figure}

\begin{figure}
\begin{center}
\includegraphics[width=13.cm]{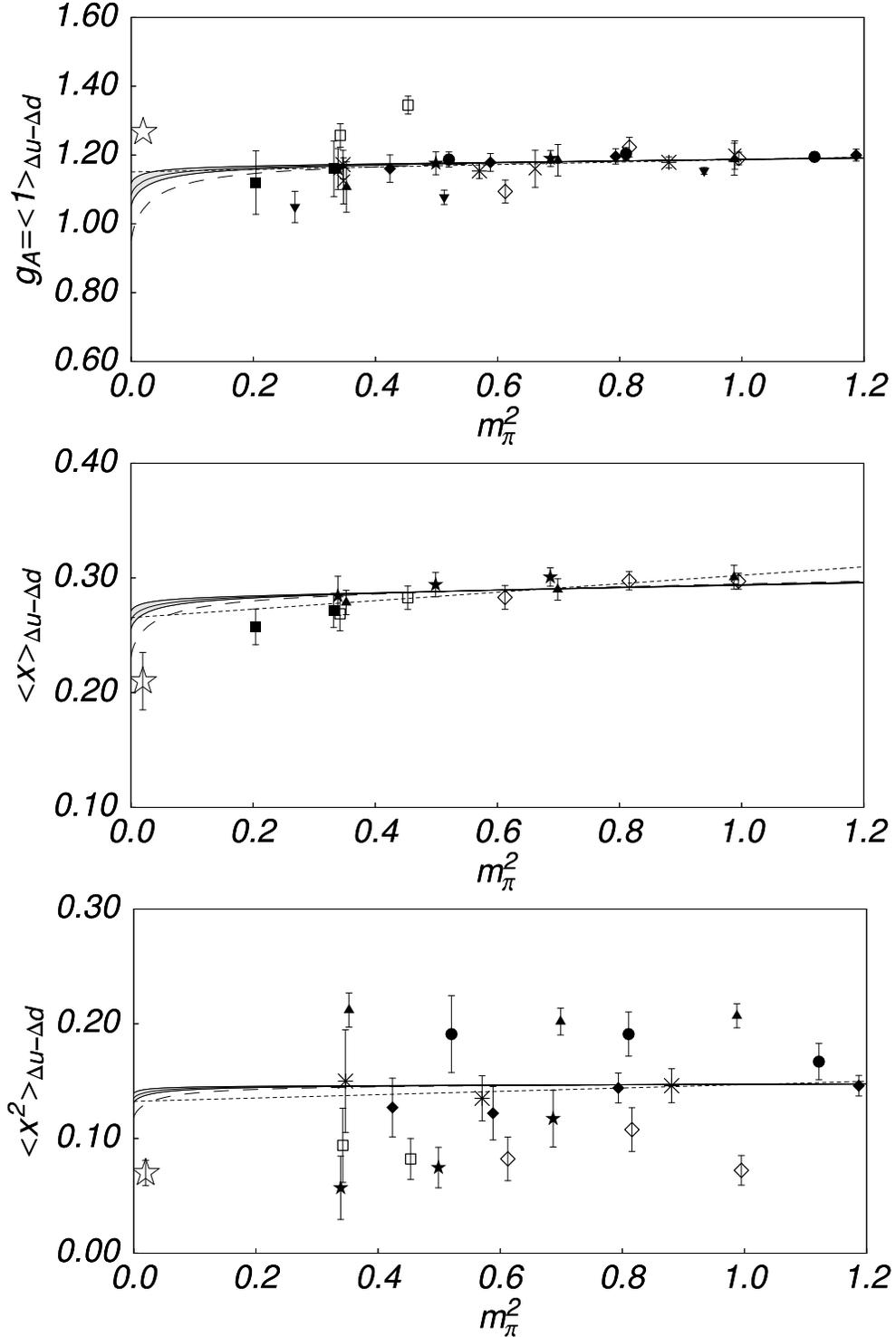}
\vspace{-0.8cm}
\caption{The lowest three moments of the helicity distribution
  $\Delta u -\Delta d$, extrapolated using a naive linear extrapolation
  (short-dashed lines) and the improved chiral extrapolation described
  in the text.
  In each panel, the long-dashed lines correspond to fits with no
  $\Delta$ and the LNA coefficient determined from $\chi$PT, while the
  solid lines are fits obtained using $g_{\pi N\Delta}/g_{\pi NN}=2$
  (upper solid curves) and $\sqrt{72/25}$ (lower solid curves).
  The lattice data are taken from the sources listed in
  Table~\protect\ref{T:data}.}
\label{F:polarized}
\end{center}
\end{figure}

\begin{figure}
\begin{center}
\includegraphics[height=14cm]{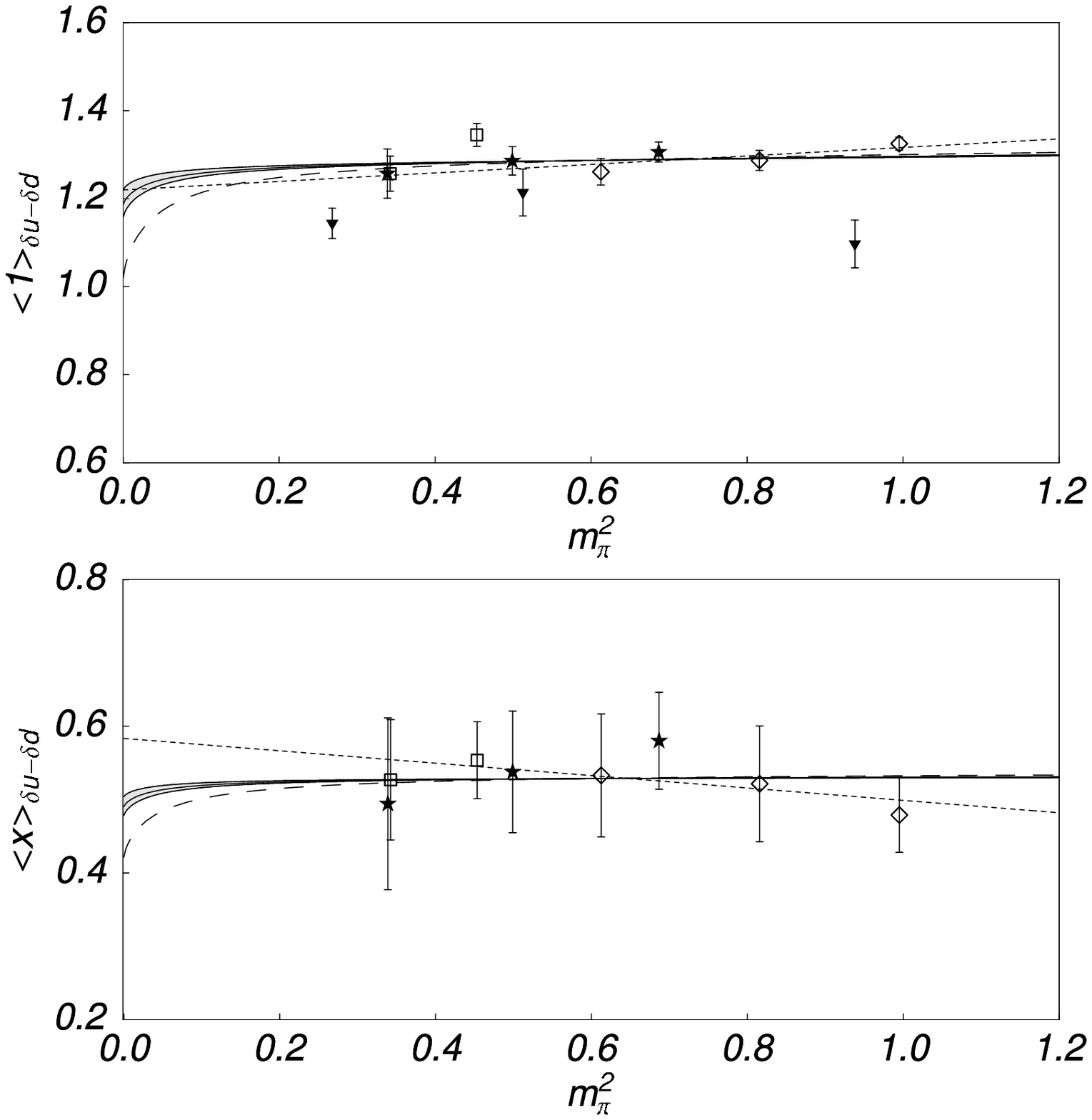}
\caption{The lowest two moments of the transversity distribution
  $\delta u -\delta d$.
  All curves are as described in Fig.~\protect\ref{F:polarized}.}
\label{F:transversity}
\end{center}
\end{figure}

Having established that the LNA formula, Eq.~(\ref{renfit}), provides a
good approximation to the full calculation, in this section we examine
the effects of extrapolating the available lattice data on the twist-2
PDF moments using the forms (\ref{xtrap}), (\ref{pxtrap}) and
(\ref{txtrap}), with the LNA coefficients determined in the limit
$\Delta M \to 0$.
Rather than show the moments versus the scale and renormalization scheme
dependent quark mass, Figs.~\ref{F:unpolarized}, \ref{F:polarized} and
\ref{F:transversity} give the moments of the unpolarized, helicity and
transversity distributions, respectively, as a function of the pion mass
squared.
The data have been extrapolated using a naive linear extrapolation
(short-dashed lines), as well as the improved chiral extrapolations
with the LNA chiral coefficients and values of $\mu$ given in
Table~\ref{fittab}, with $m_{b,n}$ fixed at 5 GeV \cite{Detmold:2001dv}.

For the spin-dependent moments, four curves are shown in
Figs.~\ref{F:polarized} and \ref{F:transversity}: the long-dashed curves
correspond to ignoring $\Delta$ intermediate states ($g_{\pi N\Delta}=0$),
while the central solid lines in each panel of the figures include the
effects of the $\Delta$ with a coupling ratio
$g_{\pi N\Delta}/g_{\pi NN}=1.85$ (average of SU(6) and phenomenological
values) and the central value for the Weinberg-Tomozawa coefficient,
$C_{WT}=0.255$.  The upper and lower
solid lines correspond to $g_{\pi N\Delta}/g_{\pi NN}=2$,
$C_{WT}=0.30$ and $g_{\pi N\Delta}/g_{\pi NN}=\sqrt{72/25}$,
$C_{WT}=0.21$, respectively.  Because the effect of the $\Delta$ is
almost negligible for the unpolarized moments, these curves are all
essentially collinear, and for clarity only one is shown in
Fig.~\ref{F:unpolarized}.
The extrapolated values are shown in Table~\ref{Tfit}, along with the
associated errors (which are described in the Appendix) and the
experimental values for the unpolarized and helicity moments
\cite{CTEQ,MRS,GRV,BB} (there are currently no data for the transversity
moments).
Note that even though there is no curvature expected for large $m_\pi$,
the slopes of the linear and LNA fits differ slightly at large $m_\pi$
values due to the constraints of the heavy quark limit incorporated into
the forms (\ref{xtrap}), (\ref{pxtrap}) and (\ref{txtrap}).
\begin{table}
  \begin{ruledtabular} \begin{tabular}{ccccccc}
    Moment&\multicolumn{2}{c}{Value}& \multicolumn{3}{c}{Extrapolation
    errors} & $a_n,\Delta a_n,\delta a_n$ \\ & Experimental &
    Extrapolated & Statistical & $\Delta$, WT states & $\mu$ & \\
    \hline $\langle x \rangle_{u- d}$ & 0.145(4) & 0.176 & 0.012 &
    0.0008 & 0.022 & 0.141\\ $\langle x^2 \rangle_{u- d}$ & 0.054(1)
    &0.054 & 0.015 & 0.0003 & 0.007 & 0.044 \\ $\langle x^3
    \rangle_{u- d}$ & 0.022(1) & 0.024 & 0.008 & 0.0001 & 0.003 &
    0.019\\ $\langle 1 \rangle_{\Delta u- \Delta d}$ & 1.267(4) &
    1.124 & 0.045 & 0.020 & 0.022 & 1.084 \\ $\langle x
    \rangle_{\Delta u- \Delta d}$ & 0.210(25) & 0.273 & 0.015 & 0.005
    & 0.005 & 0.262 \\ $\langle x^2 \rangle_{\Delta u- \Delta d}$ &
    0.070(11) & 0.140 & 0.035 & 0.003 & 0.003 & 0.135\\ $\langle 1
    \rangle_{\delta u- \delta d}$ & --- & 1.224 & 0.057 & 0.019 &
    0.025 & 1.187 \\ $\langle x \rangle_{\delta u- \delta d}$ & --- &
    0.506 & 0.089 & 0.008 & 0.010 & 0.490 \\ \end{tabular}
    \end{ruledtabular} \caption{\label{Tfit} Values of the
    unpolarized, helicity and transversity moments, extrapolated to
    the physical pion mass using Eqs.~(\ref{xtrap}), (\ref{pxtrap})
    and (\ref{txtrap}) and the parameters in
    Table~\protect\ref{fittab}.
    The experimental and systematic errors listed here are described in
    the Appendix.
    For comparison, experimental values of the
    moments where known (unpolarised values from moments of distributions of
    Refs.~\cite{CTEQ,MRS,GRV}, helicity moments from Ref.~\cite{BB} 
    in scenario I (NLO)) and the best fit parameters
    ($a_n$, $\Delta a_n$, $\delta a_n$) are also listed.}
\end{table}

With respect to the moments of the unpolarized PDFs, this analysis
confirms our earlier finding that it is essential to incorporate the
correct non-analytic behavior into the chiral extrapolation.
When this is done, there is good agreement between the extrapolated
moments at the physical pion mass and the corresponding experimental
data.
On the other hand, for the polarized PDFs we have the surprising result
that once the $\Delta$ resonance is included, the effect of the
non-analytic behavior is strongly suppressed, and a naive linear
extrapolation of the moments provides quite a good approximation to the
more accurate form. 

In the case of the axial charge (the $n=0$ moment of $\Delta u-\Delta d$),
the extrapolated value lies some 10\% below the experimental value, with
an error of around 5\%.
However, $g_A$ appears to be particularly sensitive to finite volume
corrections, with larger lattices tending to give larger values of $g_A$.
Furthermore, there is some sensitivity to the choice of action ---
simulations with domain wall fermions (DWF), which satisfy exact chiral
symmetry, are found to give larger values than those with Wilson
fermions~\cite{Sasaki:2001th}.
As almost all of the currently available lattice data are obtained from
very small lattices ($L \sim 1.6$~fm),
we consider the current level of agreement quite satisfactory.

Additionally, there is some uncertainty arising from the inclusion of
the heavy quark limit in our fits; if this constraint is omitted, the
large $m_\pi^2$ behavior of our fits coincides with the linear fits
that are shown as one would expect. For $\langle x\rangle_{\delta u
  -\delta d}$ ($\langle x\rangle_{\Delta u -\Delta d}$), a fit
ignoring the heavy quark limit gives a physical value of 0.559
(0.257) rather than 0.506 (0.273) as given in Table~\ref{Tfit} (with a
smaller effect in the other moments).

The uncertainty in the experimental determination of the higher
moments of the spin-dependent PDFs is considerably larger, and from
the current data one would have to conclude from
Fig.~\ref{F:polarized} that the level of agreement between experiment
and the extrapolated moments is acceptable.  Clearly the scatter
in the lattice data for the second moment means that at present we
cannot have much confidence in the predicted value.  We do note, in
addition, that this is one case where there is a tendency for the full
QCD points to lie somewhat below the quenched QCD results.  It is
obviously of some importance that this issue be resolved in future
lattice simulations.

\section{Conclusion}
\label{S:conclusion}

The insights into non-perturbative hadron structure offered by the
study of parton distribution functions makes this an extremely
interesting research challenge.  It is made even more important and
timely by the tremendous new experimental possibilities opened by
facilities such as HERMES, COMPASS, RHIC-Spin and Jefferson Lab.
Lattice QCD offers the only practical method to actually calculate
hadron properties within non-perturbative QCD, and it is therefore
vital to test how well it describes existing data.  Because current
limitations on computer speed restrict lattice simulations to quark
masses that are roughly a factor of 6 too large, one must be able to
reliably extrapolate the lattice data to the physical quark (or pion)
mass in order to compare with experiment.

Traditionally such extrapolations have been made using a naive linear
extrapolation as a function of $m_\pi^2$ (or quark mass).
In Ref.~\cite{Detmold:2001jb}, Detmold {\it et al.} showed that it was
essential to include the leading non-analytic behavior of chiral
perturbation theory in this extrapolation procedure.
Only then were the existing lattice data for the moments of the
unpolarized parton distribution functions in agreement with the
experimental moments.
Here we have confirmed this conclusion by calculating the next-to-leading
non-analytic behavior within a chiral quark model, including the
$\Delta$-isobar, and showing that it led to precisely the same conclusion. 

We have also investigated the variation of the moments of the polarized
parton distributions to next-to-leading order.
In this case the inclusion of the $\Delta$-isobar makes a dramatic
difference.
Indeed, once the $\Delta$ is included, the helicity and transversity
moments show little or no curvature as the chiral limit is approached
and a naive linear extrapolation formula is reasonably accurate.
In case a more accurate extrapolation procedure is desired, we propose
convenient formulae which suitably build in the non-analytic behavior in
both the unpolarized and the polarized cases.
The value of $g_A$ extracted from the extrapolation procedure at the
physical pion mass is within 10\% of the experimental value.
Given the sensitivity of this quantity to lattice volume (current
simulations use quite small lattices) and quark action (domain wall
fermions tend to give a larger value of $g_A$ than Wilson fermions),
we consider this a very satisfactory result. 
We look forward with great anticipation to the next generation of lattice
simulations of parton distribution functions at smaller quark masses and
on larger volumes.

\section*{Acknowledgements}

We would like to thank D.~Leinweber, M.~Oettel, S.~Ohta, S.~Sasaki, 
G.~Schierholz and R.~Young for helpful discussions.
This work was supported by the Australian Research Council, the
University of Adelaide and the U.S. Department of Energy contract
\mbox{DE-AC05-84ER40150}, under which the Southeastern Universities
Research Association (SURA) operates the Thomas Jefferson National
Accelerator Facility (Jefferson Lab).

\appendix
\section{Statistical and systematic errors}

In this Appendix, we describe the estimates of the statistical and
systematic errors in our fits that are presented in Table~\ref{Tfit}.

To determine an estimate of the error associated with the statistical
uncertainty of the lattice data, we use the estimated standard
deviation. For data, $f_i$, and weights, $\omega_i$, given at abscissae
$x_i$ ($i=1,\ldots,n$), the estimated standard deviation of a fitting
form $f(x ;\vec\alpha)$ with parameters $\vec\alpha$ is:
\begin{equation}
\sigma_0 = \sqrt{\frac{1}{n-1}\sum_{i=1}^n \omega_i
\left(f_i-f(x_i ;\vec\alpha_0)\right)^2}\ ,
\end{equation}
where $\vec\alpha_0$ are the best fit parameters. The statistical
errors assigned to the fits are then determined by varying the fit
parameters ($a_n$, $\Delta a_n$, $\delta a_n$) from their optimal
values (given in the right-most column of Table~\ref{Tfit}) to obtain
an increase of unity in the standard deviation.

In order to estimate the systematic errors arising from the form of
our fits, we first consider the uncertainty in the values of $g_{\pi
N\Delta}/g_{\pi NN}$ and $C_{WT}$, taking half the difference between
the physical values of the moments obtained with $g_{\pi
N\Delta}/g_{\pi NN}=2$, $C_{WT}=0.30$ and $g_{\pi N\Delta}/g_{\pi
NN}=\sqrt{72/25}$, $C_{WT}=0.21$. We also consider the uncertainty in
the fit parameter $\mu$ by taking half the difference between the
physical moments obtained with $\mu$ 20\% above and below the fits
obtained in Table~\ref{fittab}. The resulting systematic uncertainties
are listed in Table~\ref{Tfit}.

\bibliographystyle{h-physrev}

\end{document}